\documentclass[acmlarge,nonacm]{acmart}
\makeatletter
\newcommand{\myconfshort}{\acmConference@shortname}
\newcommand{\myconffull}{\acmConference@name}
\newcommand{\myconfdate}{\acmConference@date}
\newcommand{\myconfloc}{\acmConference@venue}
\AtBeginDocument{
  \fancypagestyle{firstpagestyle}{
    \fancyhead{}%
    \fancyfoot[C]{}%
  }
  \fancyhf{}
  \fancyhead[LO]{\@headfootfont\shorttitle}%
  \fancyhead[RE]{\@headfootfont\@shortauthors}%
  \fancyhead[LE]{\@headfootfont\footnotesize \myconfshort, \myconfdate, \myconfloc}%
  \fancyhead[RO]{\@headfootfont\footnotesize \myconfshort, \myconfdate, \myconfloc}%
  \fancyfoot[C]{}%
}
\makeatother
\acmBooktitle{\conffull\@ (\confshort), \confdate, \confloc}
\usepackage{tikz}
\usepackage{booktabs}
\usetikzlibrary{shapes.geometric, arrows.meta, positioning, fit, backgrounds, positioning, shapes.symbols, calc}
\usepackage{soul}
\usepackage{indentfirst}
\usepackage{longtable}
\usepackage{booktabs}
\usepackage{subcaption}
\usepackage{pdfpages}
\usepackage{cancel}
\usepackage{xcolor}

\newcommand{\new}[1]{{#1}}

\definecolor{DeepGreen}{HTML}{008000} 
\definecolor{CrimsonRed}{HTML}{DC143C} 
\definecolor{SteelGrey}{HTML}{707070}  

\newcommand{\adn}[3]{%
    (\textcolor{DeepGreen}{A: #1\%}; %
     \textcolor{CrimsonRed}{D: #2\%}; %
     \textcolor{SteelGrey}{N: #3\%})%
}
\AtBeginDocument{%
  }

\setcopyright{none}
\acmYear{2026}
\setcopyright{rightsretained}
\acmConference[FAccT '26]{The 2026 ACM Conference on Fairness, Accountability, and Transparency}{June 25--28, 2026}{Montréal, Canada}
\acmBooktitle{The 2026 ACM Conference on Fairness, Accountability, and Transparency (FAccT '26), June 25--28, 2026, Montréal, Canada (forthcoming)}
\begin{document}

\title[Taking Stock at FAccT]{``Taking Stock at FAccT'': Using Participatory Design to Co-Create a Vision for the Fairness, Accountability and Transparency Community}

\author{Shiran Dudy}
\authornote{All authors contributed equally to this research.}
\affiliation{
  \institution{Northeastern University}
  \city{Boston}
  \country{United States}
}
\orcid{0000-0002-7569-5922}
\email{s.dudy@northeastern.edu}

\author{Jan Simson}
\authornotemark[1]
\affiliation{%
  \institution{LMU Munich}
  \city{Munich}
  \country{Germany}}
\affiliation{%
  \institution{Munich Center for Machine Learning (MCML)}
  \city{Munich}
  \country{Germany}}
\affiliation{%
  \institution{University of Mannheim}
  \city{Mannheim}
  \country{Germany}}
\orcid{0000-0002-9406-7761}
\email{jan.simson@lmu.de}

\author{Yanan Long}
\authornotemark[1]
\authornote{Work began while at the University of Chicago}
\affiliation{%
  \institution{StickFlux Labs \& University of Chicago}
  \country{United States}
}
\orcid{0000-0002-7171-0443}
\email{ylong@uchicago.edu}
\email{yanan.long439@gmail.com}

\renewcommand{\shortauthors}{Shiran Dudy*, Jan Simson*, and Yanan Long*}

\begin{abstract}
As a relatively new forum, ACM FAccT has become a key space for activists and scholars to critically examine emerging AI and ML technologies. It brings together academics, civil society members, and government representatives from diverse fields to explore the broader societal impacts of both deployed and proposed technologies.
\new{We report a large-scale participatory design (PD) process for reflexive conference governance, which combined an in-person CRAFT session, an asynchronous Polis poll and the synthesis of a governance-facing report for the FAccT leadership. Participants shaped the substantive agenda by authoring seed statements, adding new statements and making patterns of agreement, disagreement and uncertainty made visible through voting.}
Our endeavors represent \new{one of the} the first \new{instances of applying} PD to a venue that critically interrogates the societal impacts of AI, fostering a niche in which critical scholars are free to voice their concerns. Finally, this work advances large-scale PD theory by providing an effective case study of a co-design paradigm that can readily scale temporally and epistemologically.
\end{abstract}




\maketitle



\section{Introduction}
Rapid developments in AI systems have \new{intensified} concerns about their under-examined societal impacts. To this end, new venues such as the ACM Conference on Fairness, Accountability, and Transparency (FAccT), the AAAI/ACM conference on AI, Ethics and Society (AIES) and the ACM conference on Equity and Access in Algorithms, Mechanisms, and Optimization (EAAMO) have been created in response to the need of critically interrogating these AI systems drawing from perspectives of diverse stakeholder groups and communities. This \new{development} dovetails with the participatory turn in AI design \cite{delgado2023participatory}, which accentuates the urgent need to broaden the sources of input into decision-making and design choices at all stages of AI system development.
\new{P}articipatory design (PD), which arose from Scandinavian workplace democracy, has frequently been invoked in projects seeking to democratize the design process (cf.\ \cite{small2023opportunities}). \new{Meanwhile, participatory language can mask participation washing, extractive practices, or weak forms of consultation} \cite{birhane2022power,sloane2022participation,arnstein1969ladder}.
%
%
Therefore, a key \new{direction} in recent PD theory is to scale-up the scope of its praxis to encompass broader geographical locations, marginalized communities and pluralistic epistemologies. In the case of participatory AI, in particular, the hyper-scale nature of contemporary AI systems \new{--- gathering adoption at unprecedented speeds ---} necessitates the inclusion of widespread participation \cite{young2024participation}.
\new{Recent PD scholarship further argues that scaling should not be treated as a single variable and that contemporary cases should be explicit about whether participation is direct or indirect, early or late, and singular or multi-technique \cite{wacnik_participatory_2025,bossen_scaling_2025}.}

\new{In this work, we present a case study of large-scale participatory design for reflexive conference governance. The process combined an in-person CRAFT session, an asynchronous Polis process and the synthesis of a governance-facing report in order to examine how the FAccT conference and collective understand the conference's present condition and future direction.} Our contributions are as follows:
\begin{enumerate}
    \item \new{One of the first case studies of large-scale PD for reflexive governance in an AI-related conference;}
    \item \new{A list of clear and actionable takeaways of how the FAccT conference might change based on an analysis of the views, expectations and aspirations of the community;}
    \item \new{An account of how participant-authored statements and voting patterns were translated into a governance-facing report for FAccT leadership;}
    \item \new{An analysis of the affordances and limits of infrastructure-mediated participation for conference governance, especially its strengths for reach, relative anonymity, and disagreement mapping and its weaknesses for iterative co-elaboration and shared control over implementation.}
\end{enumerate}

\section{Background}
\subsection{Genealogy of FAccT}
FAccT started as the \emph{Fairness, Accountability, and Transparency in Machine Learning} (FAT/ML) workshop series\footnote{\url{https://www.fatml.org/}} organized at NeurIPS (2014), ICML (2015, 2018) and KDD (2017), becoming a standalone ACM conference in 2018. Initially concerned with fairness in machine learning (ML) models\footnote{\url{https://www.fatml.org/schedule/2014/page/scope-2014}}, FAccT has seen its envisioned scope broadened to sociotechnical governance, law and policy, philosophy and normative foundations\footnote{\url{https://facctconference.org/2026/cfp.html}}, with a notable emphasis on interdisciplinary research, gathering scholars and practitioners from diverse intellectual traditions and positionalities. \new{FAccT has also become an object of reflexive scrutiny: Prior work has examined its topics, values, as well as geographic and epistemological biases, identifying gaps between its stated ambitions and what it has been able to achieve in practice \cite{laufer2022four,septiandri2023weird,schroeder2025disclosure}.} 
%

\subsection{Participatory Design in Conference Co-Design}
Despite growing application of PD to AI systems (e.g.\ \cite{huang2024collective,tseng2025ownership}), there has been a notable lack of applications of PD to the governance and organizational structures of major academic conferences in the AI ethics/responsible AI and HCI. One exception is \textit{Queer in AI}, which serves as a notable case study of community-led participatory design that has successfully transformed multiple AI conferences \cite{queerinai2023queer}. The organization has helped conferences including NeurIPS, ICML and ACL revise registration forms and diversity surveys to be more queer-inclusive, shaped author guidelines for accessibility and inclusivity, worked with conference teams to protect queer participants' privacy, and instituted effective name-change processes.

\subsection{Participatory Design in Governance}
By contrast, PD has seen many fruitful applications in public policy. One notable example is \texttt{vTaiwan}\footnote{\url{https://info.vtaiwan.tw/}}, which enabled citizens, civil-society organisations, experts, and elected representatives to discuss proposed laws via its website, as well as in face-to-face meetings and hackathons. This platform was build to replace online spaces for debate as they were engineered to capture attention, rather than encourage a productive civil discourse. vTaiwan's goal was to help policy makers improve decision making and give legitimacy to legislation through public participation. For instance, in 2015, the ride-sharing company Uber was trying to enter the Taiwanese market and was encountering resistance from local taxi drivers, leading to a flurry of debate. Employing vTaiwan, residents found paths for agreement as they coalesced around statements like ``the government should set up a fair regulatory regime,'' and ``private passenger vehicles should be registered.'' The two camps came to agree that there should be a level playing field between Uber and taxi firms. It was a scenario which protected customers while also creating more competition. In the end, the Taiwanese government would create new regulations that were informed by the consensus found on vTaiwan.

\subsection{Large-Scale Participatory Design}
Although PD is often portrayed as a methodology that fits best within small, co-located projects in workplaces or neighbourhoods, PD has always carried an ambition to intervene in larger socio-technical systems and to transform institutional arrangements, not merely to fine-tune individual artifacts. Recent systematic reviews confirm that the bulk of published PD work still reports relatively bounded case studies, frequently focused on intangible systems and services, even as the approach is increasingly invoked in discussions of large-scale infrastructures, public services and transnational issues \cite{wacnik_participatory_2025}. Within this context, talk of ``scaling'' PD has proliferated, but the term itself is used in heterogeneous and often underspecified ways. At the same time, however, it has been noted that the vast majority of works in large-scale PD are situated in the software domain \cite{zahlsen_challenges_2022}.

\section{Methods}
\begin{figure}[htbp]
\centering
    \resizebox{0.99\textwidth}{!}{%
        \begin{tikzpicture}[
            node distance=1.5cm and 2cm,
            >={Stealth[round]},
            box/.style={rectangle, draw, rounded corners, minimum width=2cm, minimum height=1cm, align=center, fill=white},
            phase/.style={rectangle, rounded corners, draw=gray!60, fill=gray!10, inner sep=6pt},
            arrow/.style={->, thick, >=Stealth}
        ]
        \node[box] (seed) {Seed\\Statements};
        \node[box, left=1.5cm of seed, yshift=0.8cm] (researchers) {Researchers};
        \node[box, left=1.5cm of seed, yshift=-0.8cm] (participants1) {Participants};
        
        \draw[arrow] (researchers.east) -- node[above, font=\small, yshift=0.2cm, xshift=0.4cm] {1) write prefix} (seed.west);
        \draw[arrow] (participants1.east) -- node[below, font=\small, yshift=-0.2cm, xshift=0.85cm] {2) complete statement} (seed.west);
        
        \begin{scope}[on background layer]
            \node[phase, fit=(researchers)(participants1)(seed), label={[anchor=south]north:\textbf{Phase 1: Seed Statement Generation}}] (phase1) {};
        \end{scope}
        
        \node[box, right=2.5cm of seed] (polis) {Polis\\Platform};
        \node[box, right=1.5cm of polis, yshift=0.8cm] (vote) {Statements};
        \node[box, right=1.5cm of polis, yshift=-0.8cm] (submit) {New\\Statements};
        
        \draw[arrow] (seed.east) -- node[above, font=\small] {3) register to} (polis.west);
        \draw[arrow] (polis.east) -- node[above, font=\small, yshift=0.1cm, xshift=-0.1cm] {4) vote} (vote.west);
        \draw[arrow] (polis.east) -- node[below, font=\small, yshift=-0.1cm, xshift=-0.1cm] {4) submit} (submit.west);
    
        \draw[arrow] (submit.north) -- node[left, font=\small] {5) reflect} (vote.south);
        
        \begin{scope}[on background layer]
            \node[phase, fit=(polis)(vote)(submit), label={[anchor=south]north:\textbf{Phase 2: Polis Engagement}}] (phase2) {};
        \end{scope}
    
        \node[box, minimum height=1.6cm, right=2.5cm of $(vote.east)!0.5!(submit.east)$] (report) {Leadership- and\\Community-facing\\Report};
    
        \draw[arrow] ($(vote.east)!0.5!(submit.east)$) -- node[above, font=\small] {6) synthesize} (report.west);
    
        \begin{scope}[on background layer]
            \node[phase, fit=(report), inner sep=10pt, label={[anchor=south]north:\textbf{Phase 3: Report Synthesis}}] (phase3) {};
        \end{scope}
        
        \end{tikzpicture}
    }
    \caption{The Participatory design process: participants co-created seed statements with researchers, then engaged with the Polis platform to vote on existing statements and submit new ones\new{, including the option of authoring new statements to reflect on existing ones. Last, results are synthesized into a report to be shared with both conference leadership and the community.}}
    \label{fig:participatory_process}
\end{figure}
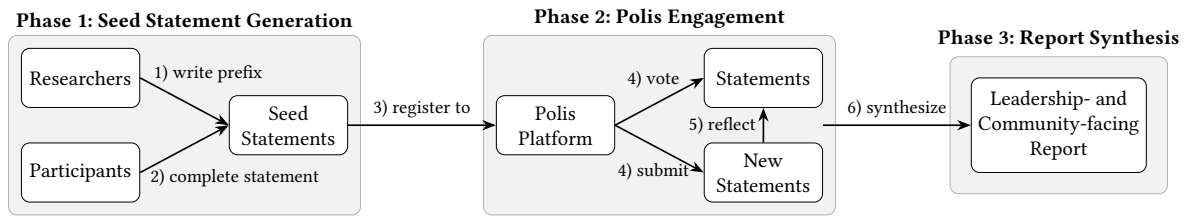

\new{\subsection{Process Design and Stages}}
The engagement process followed a \new{three-phase} structure. Phase 1 (Figure~\ref{fig:participatory_process}, left) took place as an in-person CRAFT session held on June 26, 2025 at FAccT'25 to co-create seed statements (a session that was initiated with a broad introduction to participatory design). \new{We introduced the session as the agenda-setting stage of a participatory design process focused on how participants understand FAccT today and what they want it to become. To scaffold that work, we provided a set of prefixes: ``FAccT is...'', ``FAccT is not...'', ``FAccT should...'', ``FAccT should not...'', ``FAccT is doing well in...'', and ``FAccT is not doing well in...''.} CRAFT session participants were asked to complete these prefixes by adding their unique perspectives, thereby co-creating complete statements. We organized participants into three groups of two, yielding a total of 33 seed statements. At the conclusion of the CRAFT session, the seed statements were registered to Polis \cite{small2021polis}, marking the beginning of Phase 2 (Figure~\ref{fig:participatory_process}, right). CRAFT participants were given early access to vote on these statements and to submit new statements of their own, and this access was later on expanded to the wider FAccT community. Unlike the seed statements, these new contributions were unconstrained by any predetermined prefix. Phase 2 extended beyond the in-person session, continuing online from June 26 to August 1, 2025. \new{Once the information gathering process was completed, we synthesized our findings into a report as shown in Phase 3, with the goal of sharing it with both FAccT's leadership and its wider community.}



\subsection{Participatory Design Tool}
Polis is a deliberative democracy platform that enables scalable online participation with low barriers to entry~\cite{small2021polis}, distinguishing itself from traditional discussion forums by prohibiting direct replies between participants and instead using voting patterns to generate visual representations of opinion distribution and consensus areas across the participant population. The platform allows participants to vote on existing statements, either agreeing, disagreeing, or passing on a statement and also to submit their own statements (statements are under $ 140 $ characters). In our study, submitted statements were unmoderated and became immediately available for voting as recommended by~\cite{small2023opportunities}. We encouraged participants to provide an email address when submitting new statements to enable follow-up communication. However, this was not a requirement. CRAFT session participants were the first to engage with the online voting process.


\subsection{Recruitment}
%
The CRAFT session and Polis platform were launched on June 26, 2025. The online participation period remained open until August 1 AoE, with most votes being cast just before the communicated deadline of July 31 AoE. During this time, we recruited additional community members through announcements on the FAccT-discuss listserv and the FAccT Slack general channel, sending two emails and three Slack posts over this period, on June 29, July 18, and then July 31. To reach potential participants, we disseminated invitations exclusively through the FAccT Slack channel and the FAccT-discuss listserv. However, this open participation model presents a limitation: we cannot assess the representativeness of the sample relative to the broader FAccT community. 

\subsection{Data Processing}\label{postprocess}



\subsubsection{Data Cleaning}
After the voting period concluded, we manually inspected all statements and removed two that were incomplete or unintelligible, yielding a total of $ 59 $ statements (Table~\ref{tab:pd_statements_labeling}). The remaining statements underwent a deduplication process: the statements were first embedded by the bi-encoder \texttt{sentence-transformers/all-MiniLM-L6-v2} and then sentence pairs with cosine similarities greater than a threshold were chosen as candidates, which were subsequently validated by the cross-encoder \texttt{cross-encoder/ms-marco-MiniLM-L6-v2} using a threshold of $ 0.667 $. Following \citeauthor{small2023opportunities}~\cite{small2023opportunities}, who has cautioned that removing statements for irrelevance or duplication risks silencing participant voices or introducing bias, we adopted a conservative approach with a high similarity threshold for the bi-encoder. We empirically tested cosine similarity thresholds and found that at $ 0.75 $, the identified pairs were not sufficiently similar to warrant removal. At $ 0.8 $, no duplicates were found, which is the threshold we adopted.



\subsubsection{Thematic Analysis}\label{theme_ana}
The thematic analysis (\S~\ref{sec:topic}) required two steps; first determining the theme labels, second, labeling the statements. Submitted Statements were fed into Claude Opus 4.5 to generate theme labels (topics), which were subsequently validated by two of the authors serving as human annotators. The list of labels as well as their definitions can be found in Table~\ref{tab:themes_def}. All statements submitted by the participants were labeled by the same annotators and by Claude Opus 4.5, Gemini Pro 3 and GPT 5.2. Each statement was allowed to be assigned to more than one theme, and the labels preferred by the majority of LLM and human annotators (including ties) was chosen as the theme label(s) for the statement for subsequent analysis.

\subsubsection{Theme-based analysis of the votes}
\label{theme_votes}
The voting analysis based on the themes in section~\ref{sec:vote_theme}, was based on the following steps. First, we extracted statements relevant to each theme along with their corresponding voting data from the Polis data. For each statement, we assessed two dimensions: degree of consensus and the participation rate. 
Consensus was categorized as strong ($ >75 \% $ agree or disagree), majority (greater than $ 50\% $ agree or disagree), high uncertainty ($ >50\% $ pass rate), or mixed opinions (all other cases). Participation rate (or vote count) was classified as high (more than 80 votes), moderate (51–80 votes), or low (50 or fewer votes). Each statement was then labeled with a summary string following this format: ``[consensus category] ($ x\% $ agree, $ y\% $ disagree, $ z\% $ pass) - [participation rate]''. The resulting theme-based CSV files were interpreted using an automated system prompted to summarize the comments and voting patterns while accounting for consensus and participation rates. One author reviewed these summaries to ensure they accurately described the voting patterns within each theme.

\subsection{\new{Report}}\label{report_sec}

\new{
A key goal of this project is to facilitate change in the FAccT conference in line with the community's ideals. To this end, we planned to create an artifact as the final output of the participatory design process. In coordination with FAccT leadership we arrived on a report as the preferred format for this artifact and shared an initial draft with the conference leadership.
This report is a preliminary artifact and part of the participatory process. We plan to share the report as well as the participatory data itself directly with the FAccT commmunity, allowing for further feedback, iteration and secondary analyses.
}

\section{Results}

\subsection{Participant Engagement}

\begin{figure}
    \centering
        \includegraphics[width=0.95\textwidth]{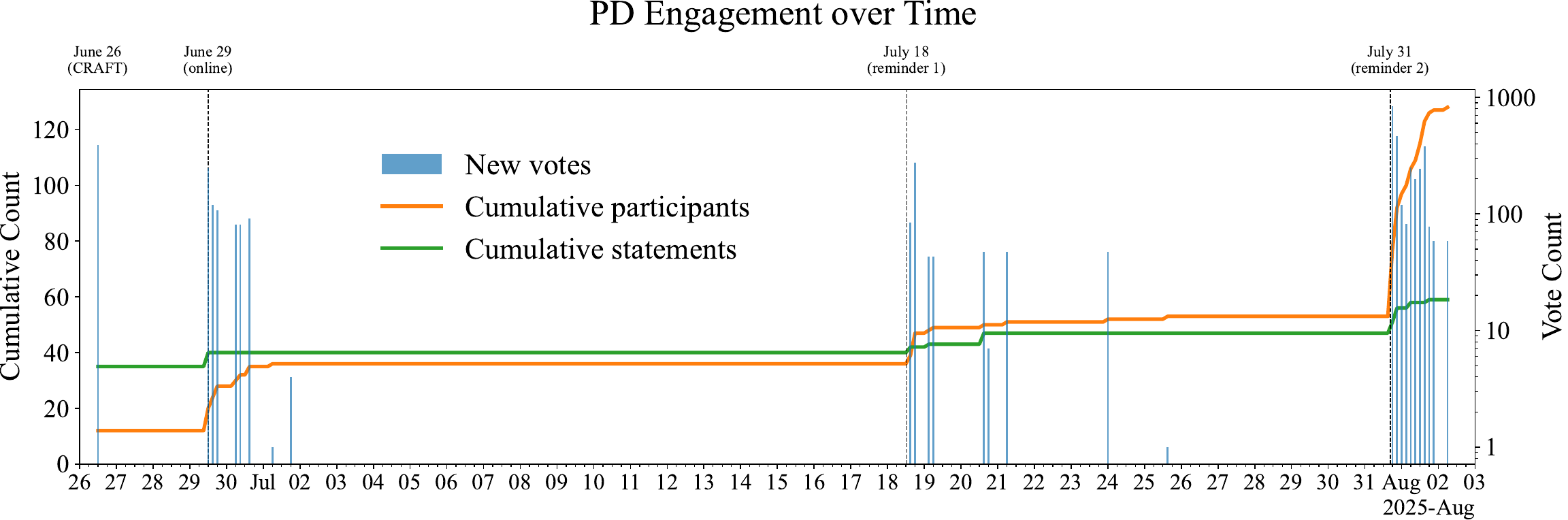}
        \caption{Participation over time as measured by the cumulative rate of new participants casting their first vote (left $y$-axis, orange line), the cumulative number of statements (left $y$-axis, green line), and the number of votes cast per 3-hour windows (right $y$-axis, blue).}
        \label{fig:accu_engagement}
\end{figure}

\subsubsection{Participation over time}
We investigated when participants exhibited increased activity. To this end, Figure~\ref{fig:accu_engagement} presents engagement metrics over time. The cumulative number of participants over time are plotted with new participants identified by their first recorded vote (left $ y $-axis, orange line), alongside the cumulative number of statements that have been submitted (left $y$-axis, green line), and the number of new votes on a given calendar day (right $ y $-axis, blue bars). Vertical gray lines indicate key milestones: the CRAFT session, the initial online recruitment, and two subsequent reminders. 
The process began with 33 seed statements, with two additional statements submitted on the same day. By the end of June 26, 12 participants had joined, casting a total of 391 votes (right y-axis). The online launch on June 29 generated a new wave of activity through July 1, bringing to a total of 36 participants, another round of votes totaling in 1,122 votes, and 40 statements. Following the July 18 reminder, participation increased to 53 participants with a longer voting wave, across 49 statements (including 9 newly submitted). The final reminder on July 31 resulted in 128 total participants, a total of 4,531 votes, and 61 statements (two were then discarded as mentioned above).

All three measures were found to to behave similarly over time. We find that engagement rates expressed in new voters, new statements, and votes being cast, grew over time, with higher activity around the recruitment dates, and highest activity at the final reminder that was sent just before the end of the voting period. \textit{Nudges and reminders helped increase all three forms of engagement}. To situate the number of participants, the FAccT \texttt{\#general} Slack channel currently has 1,966 members and 128 participants therefore represent a participation rate of approximately 6.5\% of this subset of the community. 

\begin{figure}
    \centering
    \begin{subfigure}[t]{0.55\textwidth}
        \centering
        \includegraphics[width=\textwidth]{figures/votes_per_voter.pdf}
        \caption{Number of votes cast per participant across all participants. Each dot corresponds to one participant, ordered by how many votes they cast. The upper horizontal green line represents the median number of votes cast ($\tilde{n}_{votes} = 40$), the lower blue line represents the average number of votes cast ($\bar{n}_{votes} = 35.5$).}
        \label{fig:voted_per_voter_sorted}
    \end{subfigure}
    \label{fig:both_plots}
    \hfill
    \begin{subfigure}[t]{0.4\textwidth}
        \centering
        \includegraphics[width=\textwidth]{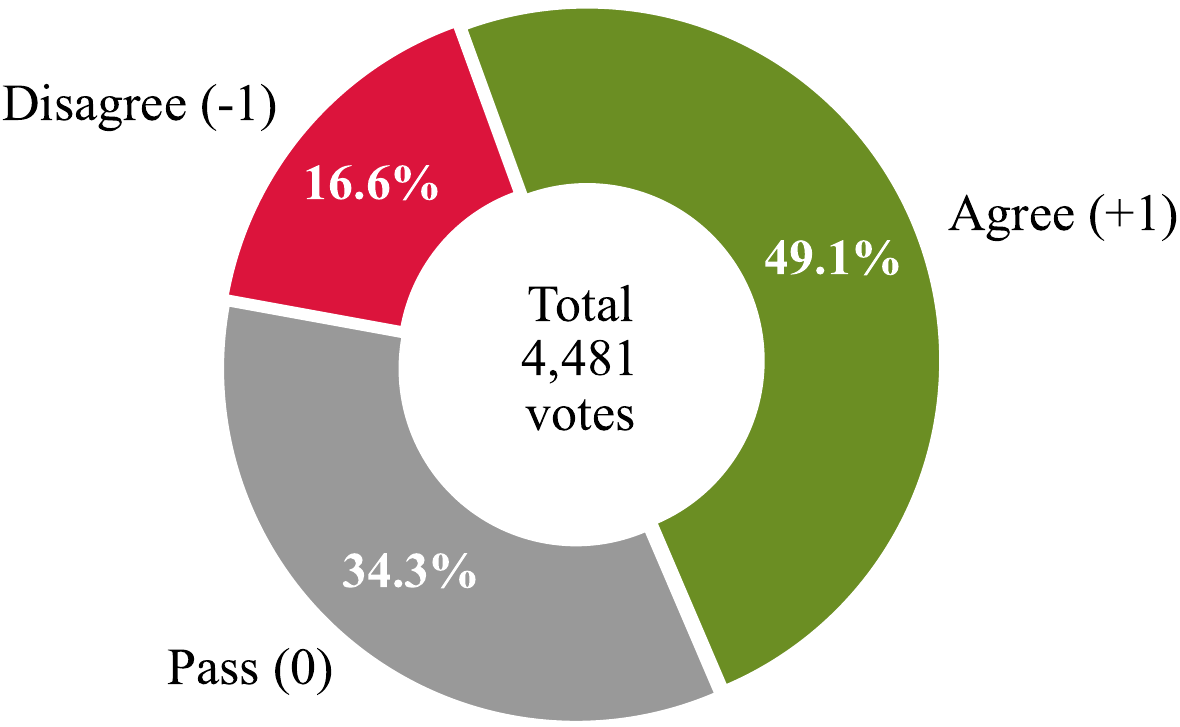}
        \caption{Overall distribution of votes across the three voting options: agree ($49.1\%$; $2,201$ votes), pass ($34.3\%$; $1,537$ votes), and disagree ($16.6\%$; $743$ votes).}
        \label{fig:vote_distrib}
    \end{subfigure}
    \hfill
    \begin{subfigure}[t]{0.48\textwidth}
        
    \end{subfigure}
    \caption{Participant voting behavior by number of votes cast (a) and overall voting distribution (b).}
\end{figure}

\subsubsection{Participation rates} We examined participant activity levels by analyzing the number of statements each participant voted on. Figure~\ref{fig:voted_per_voter_sorted} displays participants and the corresponding number of votes they cast, sorted in descending order. The median number was $\tilde{n}_{votes} = 40$, where half of participants voted on 40 statements and more, or half of the participants voted on 67\% or more of the statements (40 of 59 total). The average number of votes cast were $\bar{n}_{votes} = 35.5$ votes per participant, with approximately 80 participants (62\%) voting at or above this rate. This indicates that voting was not driven by a small number of highly active voters, but rather reflects \textit{broad engagement across the majority of participants}.

\subsubsection{Overall voting patterns}
When engaging in the voting process, participants are presented with each statement and asked to select one of three responses: agree, disagree, or pass. Figure~\ref{fig:vote_distrib} shows the overall distribution of these responses: agreement was most common at $49.1\%$ ($n = 2,201$ votes), followed by passing a statement at $34.3\%$ ($n = 1537$), and disagreement was the least common with only $16.6\%$ ($n = 743$). To summarize, \textit{participants were nearly three times more likely to agree with a statement than to disagree.}

\begin{figure}
    \centering
        \includegraphics[width=0.5\textwidth]{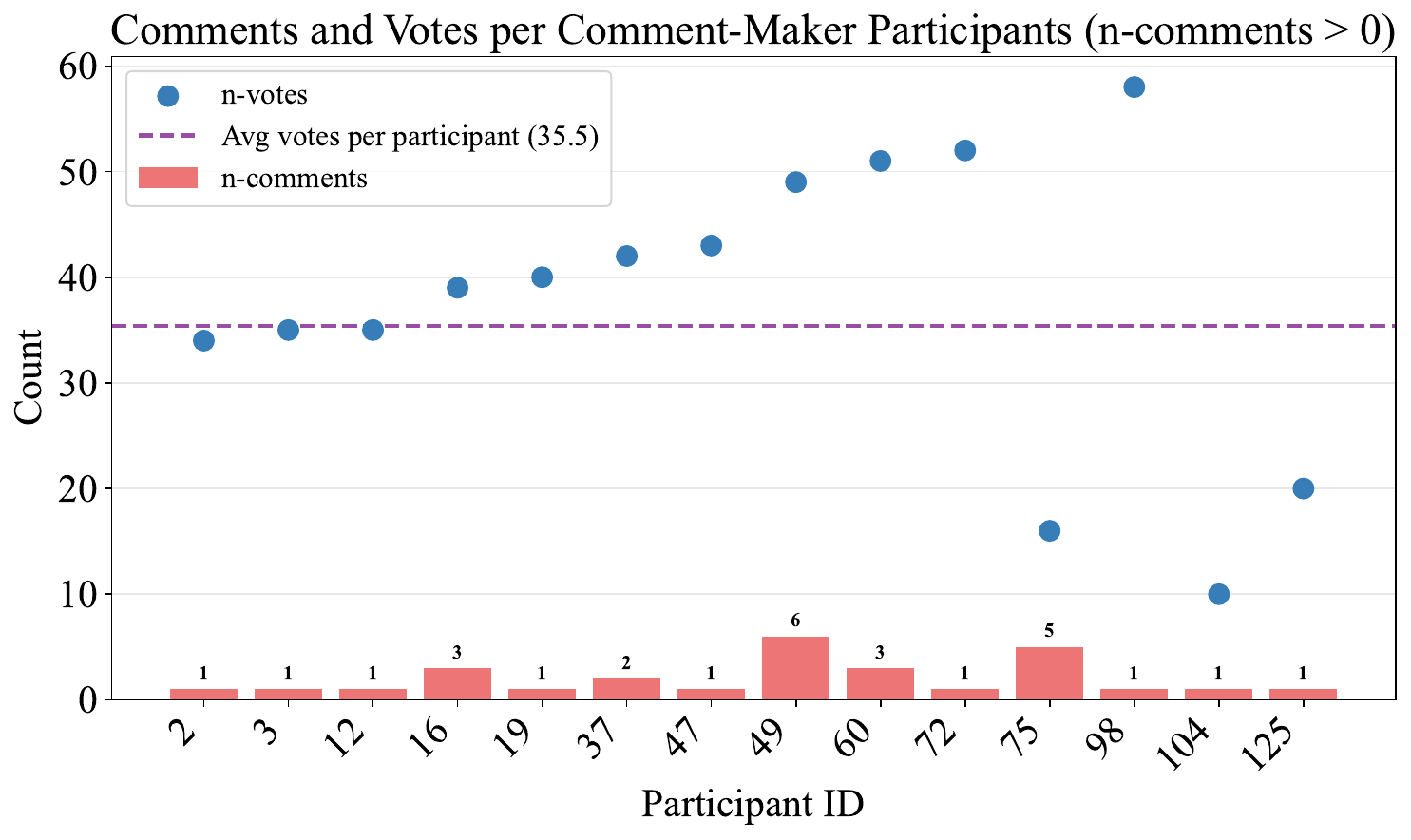}
        \caption{Engagement of participants who contributed statements. Individual statement creators are spread along the $x$-axis with bars corresponding to the number of statements contributed (in red), blue circles correspond to the number of votes cast. The dashed line indicates the average number of votes cast by participants ($\bar{n}_{votes} = 35.5$).}
        \label{fig:comment_makers}
\end{figure}
 
\subsubsection{Statement creators}
In this analysis we examined participants who opted to submit their own statements. In particular, we asked whether these participants differ in their voting behavior with respect to the average participant. 
Figure~\ref{fig:comment_makers} displays anonyomus participant IDs of all participants who submitted statements alongside the number of statements each contributed and the number of statements they voted on. This analysis is excluding seed statements, which are not attributed to individual participants. The average number of votes cast by all participants is shown as a baseline. The majority of statement creators demonstrated strong engagement, voting at or above the average. However, a small number of statement creators voted below the average, including a single participant (\#75) who submitted five statements yet voted on fewer than 20. Overall, \textit{participants who contributed statements of their own also exhibited high voting rates.}

\subsubsection{Statements' exposure}

\begin{figure}
    \centering
    \begin{subfigure}[t]{0.44\textwidth}
        \centering
        \includegraphics[width=\textwidth]{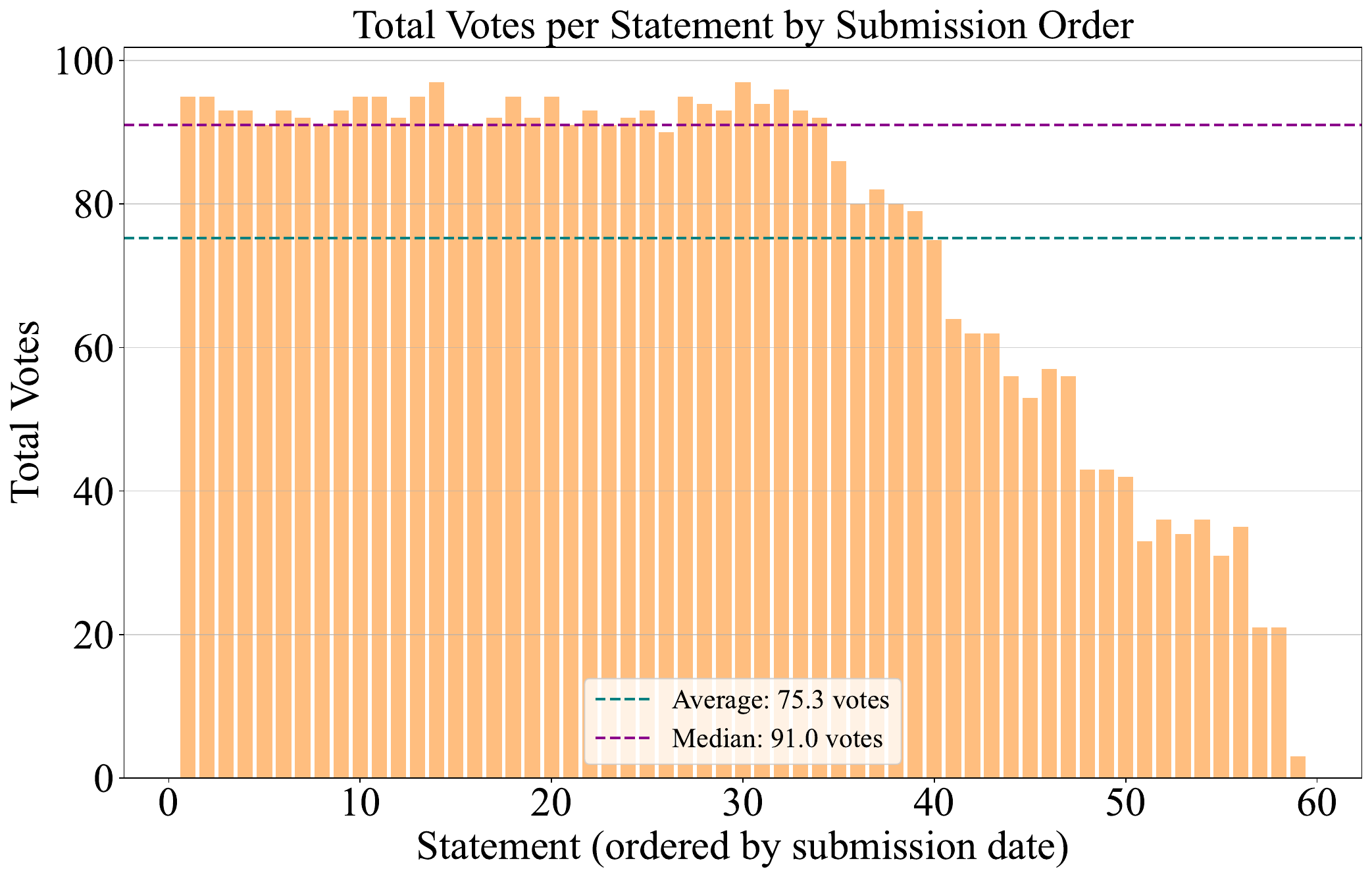}
        \caption{Analyzing the exposure of statements. Statements are ordered by their submitted date on the $x$-axis and the number of total votes they gained on the $y$-axis. The median and average are presented in horizontal lines indicating that half of the statements cast above 91 votes with an average $\bar{n}_{votes} = 73.7$ votes overall.}
        \label{fig:statement_exposure}
    \end{subfigure}
    \hfill
    \begin{subfigure}[t]{0.44\textwidth}
        \centering
        \includegraphics[width=\textwidth]{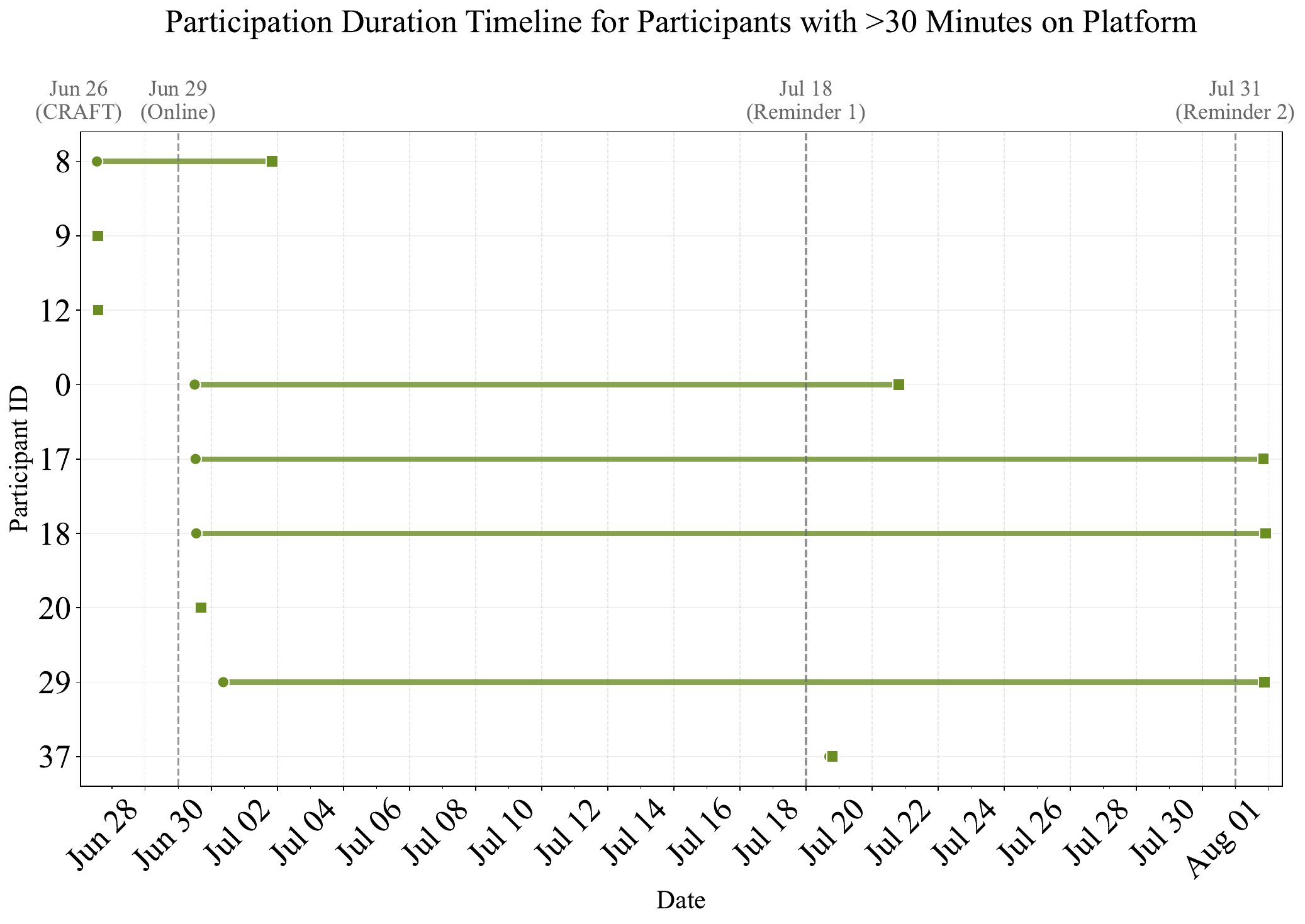}
        \caption{Participation duration for participants whose participation period (time between first and last votes cast) exceeded 30 minutes. There were $ 9 $ out of $ 128 $ participants that satisfied this criterion. Vertical lines indicate onset of CRAFT, online voting, and reminders}
        \label{fig:participation_interval}
    \end{subfigure}
    \caption{Participation engagement over time. Votes received per statement (a) and duration of participation (b).}
    \label{fig:engagement}
\end{figure}

In this analysis we sought to learn about the statement exposure. In particular, we were asking whether there is any connection between the time of the submitted statement and its total voting. 

Figure~\ref{fig:statement_exposure} presents the list of statements ordered by their date of submission, and their corresponding total number of votes on the y-axis. We find that more than half of the statements received 91 votes or higher. Moreover, a clear pattern emerges: \textit{statements submitted earlier received more votes, indicating an exposure bias favoring early contributions.}

\subsubsection{Participation duration}
As part of the onboarding process, participants were encouraged to provide their email address to receive notifications about new statements and continue voting. We examined whether participants returned to the platform after their initial engagement and prior to the study deadline. Figure~\ref{fig:engagement} shows the number of participants who returned after their first session. \textit{Only 9 out of 128 (7\%) participants returned to vote on newer statements, explaining the strong time-effect of statement exposure.}


\subsection{Statements}
\subsubsection{The conversation}
Our goal in this initiative 
was to engage community members in two types of conversations: one in which participants share how they perceive FAccT today and another aimed at exploring what community members envision for FAccT in the future. It is worth noting that the interpretation of ``FAccT'' was intentionally kept open (cf.\ \S~\ref{sec:facct_collective}).

\begin{figure}[h]
  \centering
  \includegraphics[width=0.7\textwidth]{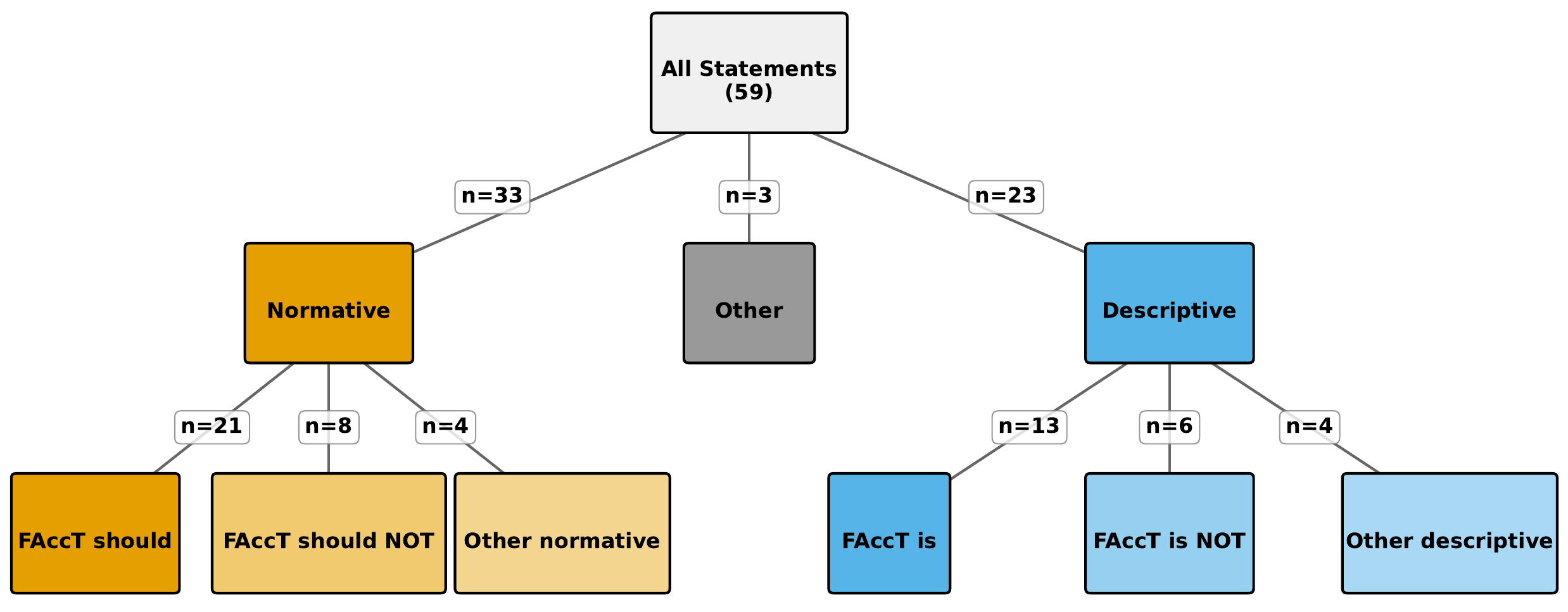}
  \caption{Classification of statements as normative and descriptive.}
  \label{fig:stat_nor_desc}
\end{figure}
To evaluate whether we successfully engaged the community in the types of conversations we intended, we analyzed the statements and categorized them into three high-level categories: \textit{Normative}, \textit{Descriptive}, and \textit{Other} (see Figure~\ref{fig:stat_nor_desc}). \textit{Normative} statements address future visions of FAccT; \textit{Descriptive} statements reflect current perceptions of FAccT; and the \textit{Other} category captures statements that did not fit either definition.
To perform this classification, we first employed a rule-based algorithm: statements with the prefixes ``FAccT should...'', ``FAccT should not...'' were labeled as \textit{Normative}, while those beginning with ``FAccT is...'', ``FAccT is not...'' were labeled as \textit{Descriptive}. Statements that did not match either of these patterns were manually analyzed based on whether they discussed FAccT in the present or future tense. The Other category contained statements that addressed neither, such as ``Polis is not a useful way to have a conversation about these issues.'' The complete Normative-Descriptive classification is provided in Table~\ref{tab:pd_statements_labeling} in the appendix.
Of the 59 statements analyzed, 33 were classified as Normative, 23 as Descriptive, and 3 as Other. This distribution indicates that we successfully facilitated conversations aligned with our intended goals.

\subsubsection{Themes} 
\label{sec:topic}
\begin{figure}[t]
  \centering
  \includegraphics[width=.9\textwidth]
  {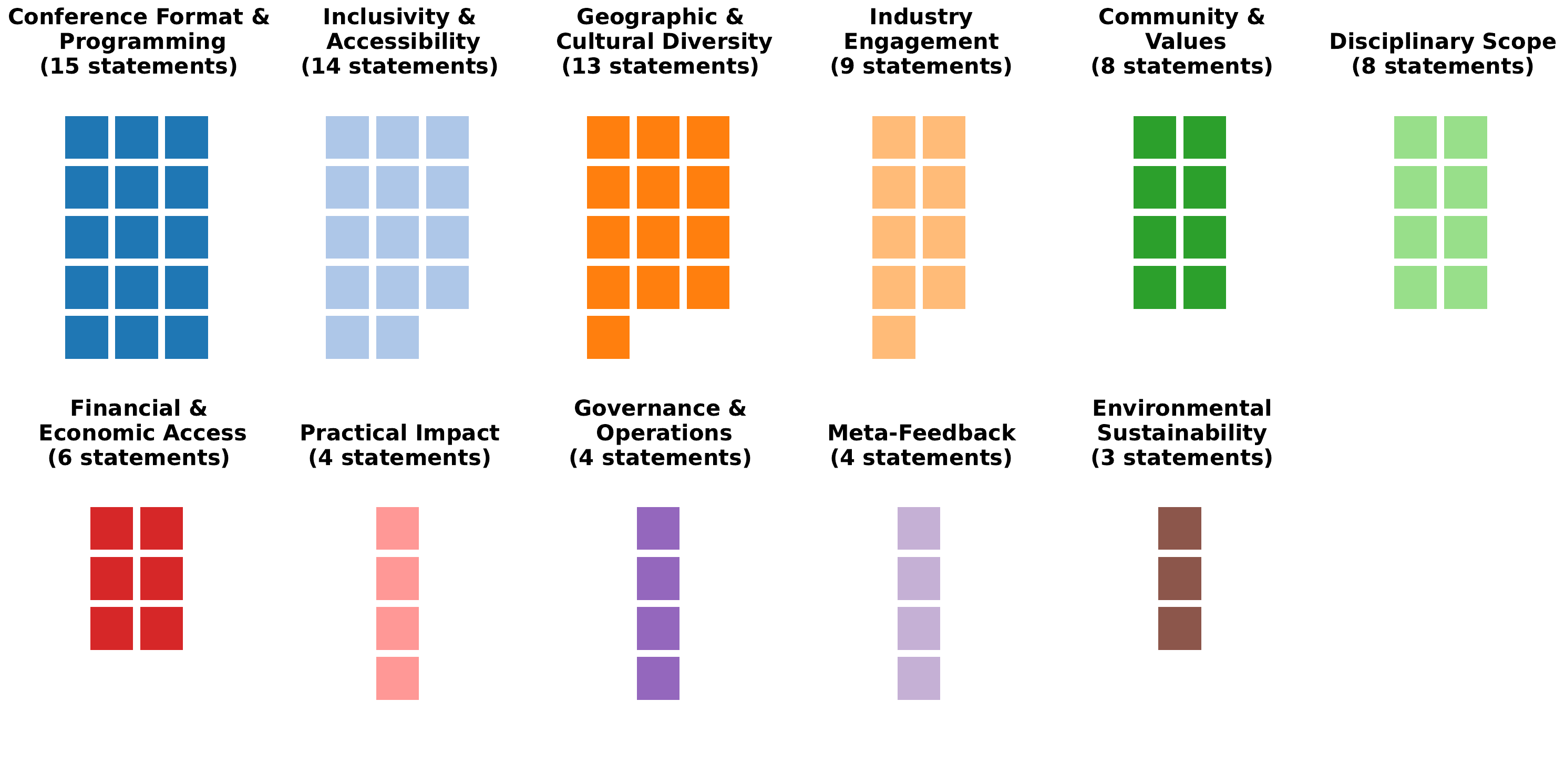}
  \caption{Distribution of all 59 statements across the 11 identified themes.}
  \label{fig:stat_themes}
\end{figure}

In this section, we present the themes emerging from the statements submitted to Polis. Identifying these themes reveals which topics are central to participants' concerns, while the relative proportions may indicate the collective importance participants attribute to each theme. We followed the process described in section \ref{theme_ana}. The count of each theme can be found in Figure~\ref{fig:stat_themes}. The overall annotator agreement measured through Krippendorff's Alpha (Jaccard distance) was 0.77, and the macro-averaged Fleiss Kappa was 0.66 both indicating moderate agreement levels. The complete list of statements and their assigned labels can be found in the appendix in Table~\ref{tab:pd_statements_labeling}. We find that the top themes emerging from the statements are Conference Format \& Programming, Inclusivity \& Accessibility, Geographic \& Cultural Diversity, Industry Engagement, and Community \& Values appearing in 15, 14, 13, 9, 8 statements respectively, where each of those themes is defined in Table~\ref{tab:themes_def}. 
\begin{table}
\small
\centering
\caption{List and definitions of the identified theme categories.}
\label{tab:themes_def}
\begin{tabular}{@{}p{4cm}p{9cm}@{}}
    \toprule
    \textbf{Theme} & \textbf{Definition} \\
    \midrule
    Inclusivity \& Accessibility & Statements about welcoming diverse participants, neurodiversity accommodations, accessibility for different groups \\
    Geographic \& Cultural Diversity & Concerns about Western/US-centrism, location choices, visa issues, local community engagement, multilingualism \\
    Disciplinary Scope & Multidisciplinary nature, CS focus, cross-disciplinary collaboration, epistemological diversity \\
    Industry Engagement & Relationship with corporations, sponsorship, industry participation vs.\ influence \\
    Financial \& Economic Access & Pricing, student fees, financial support for attendees, income-based pricing \\
    Conference Format \& Programming & Poster sessions, workshops, tracks, networking opportunities, session structure \\
    Community \& Values & Shared values, community building, solidarity structures, mutual aid \\
    Environmental Sustainability & Carbon footprint, plastic waste, vegetarian options \\
    Practical Impact & Guidance for practitioners, tangible outcomes, real-world application \\
    Governance \& Operations & Volunteer-run strucpture, organizer compensation, conference organization \\
    Meta-Feedback & Comments about this survey/process itself or general observations \\
    \bottomrule
\end{tabular}
\end{table}


\subsection{The votes} 
Building on the themes identified in Section \ref{sec:topic}, we analyzed voting patterns to surface the emerging perspectives shared by participants. Following the description in section~\ref{theme_votes} we provide a summary of the main findings, where the complete theme-based analysis of the votes can be found in the appendix.

\new{
\textit{Community \& Values.} Participants expressed strong consensus with high engagement that FAccT should be an open and welcoming community. With similarly high engagement, a majority agreed that FAccT serves as a useful space to meet future collaborators, is building a community with shared values, and is doing well at sticking to those values. 

\textit{Conference Format \& Programming.} Participants showed strong agreement with high engagement that FAccT is a useful space to meet future collaborators and should offer more opportunities to connect, such as speed dating and poster sessions. There was also majority consensus that FAccT is doing well in providing varied formats like socials and CRAFTs and that poster sessions specifically should be part of the program. 

\textit{Disciplinary Scope.} Participants showed majority consensus with high engagement that FAccT is doing well in fostering respectful exchange across interdisciplinary topics. With moderate engagement, there was also majority agreement that FAccT occupies a unique position and must remain a top venue for computing and social justice research, and that the conference should support more cross-disciplinary meetings beyond just socials. 

\textit{Environmental Sustainability.} 
The voting patterns indicate a community that values sustainability as a consideration but is divided on how far such commitments should extend into individual or collective practices at the conference.

\textit{Financial \& Economic Access.} 
The community clearly prioritizes financial accessibility and attendee support, but remains divided on the specific mechanisms, such as free student registration or limiting industry funding, for achieving these goals.

\textit{Geographic \& Cultural Diversity.} 
The community recognizes a US and Western tilt in FAccT's current orientation and broadly supports greater geographic and cultural inclusivity, but remains 
uncertain on specific mechanisms for achieving this shift.

\textit{Governance \& Operations.} Participants demonstrated high uncertainty about whether FAccT is doing well in implementing requests from previous years into future conferences, suggesting many voters felt unable to assess the conference's responsiveness to community feedback. 
There was strong consensus acknowledging that while it is easy to vote on judgments about organizing a conference, it is far more difficult to actually put together a conference that integrates all of them. 

\textit{Inclusivity \& Accessibility.} 
Overall, the community is united around aspirational values of openness and non-exclusion, and supports several practical accessibility measures, but remains divided 
on how well FAccT currently delivers on inclusivity and on some of the specific mechanisms for broadening access.

\textit{Industry Engagement.} 
The community is actively grappling with the role of industry at FAccT without having consensus on how much corporate funding or presence is appropriate.

\textit{Practical Impact.} Participants showed mixed opinions across all statements related to practical impact, with high engagement but no clear consensus on any item. There was division on whether FAccT should serve as a generator for alternative solidarity structures such as mutual aid and whether it should engage in real participatory design. 
}

Our findings suggest that the community \textit{readily identifies its ideals} and challenges, such as fostering an open and welcoming environment, promoting sustainability, ensuring financial accessibility, and addressing the conference's US and Western-centric orientation, yet there is less consensus on \textit{how} to address these issues. Certain topics elicited particularly \textit{mixed opinions}, especially regarding industry engagement, governance strategies, and whether FAccT should have practical impact beyond the conference. This participatory design process represents a first step toward understanding where the community stands on these concerns.

\subsection{\new{Report and Takeaways}}

\new{
To strike a balance between a concise report and the risk of over-aggregation of participants' statements, we opted to primarily focus on providing concise and actionable executive summaries while also making the participatory statements and votes directly accessible in the report rather than trying to convey insights solely through summarized text which may risk omitting or misrepresenting voices in the community. The report also contains short descriptions of context and methodology to accurately position it.

We opted to present statements organized alongside two axes in the report: one axis being \textit{operational statements} which are readily applicable and \textit{aspirational} statements which are more concerned with the larger mission of FAccT, the other axis being statements of \textit{high} versus \textit{low consensus} (with consensus defined as a large portion of similar votes, rather than pure agreement).  High-consensus operational statements provide clear takeaways for organizers, including ``FAccT should also run local meetups (EU/UK/Asia-Pacific) in case if attendees fail to attend due to visa reasons'' \adn{71}{0}{29}, ``FAccT should reduce sponsorship reliance to a minimum by no longer providing financial support to any attendees'' with high consensus on disagreeing with this statement \adn{3}{72}{25}, ``FAccT should have more opportunities to meet people (i.e. speeddating, poster sessions)'' \adn{68}{4}{27} to name just the three highest-consensus ones. Other concrete takeaways include tiered pricing for different countries \adn{68}{14}{19}, inclusion of wider formats at the conference [posters \adn{64}{12}{24}; workshops etc. \adn{58}{5}{37}], reduction of plastic waste \adn{64}{13}{23} and taking visa barriers into account when choosing a location for the conference (different statements highlight different angles of the issue). Low consensus statements highlight areas which will require further discussion within the community and often center around issues of (corporate) funding.

}

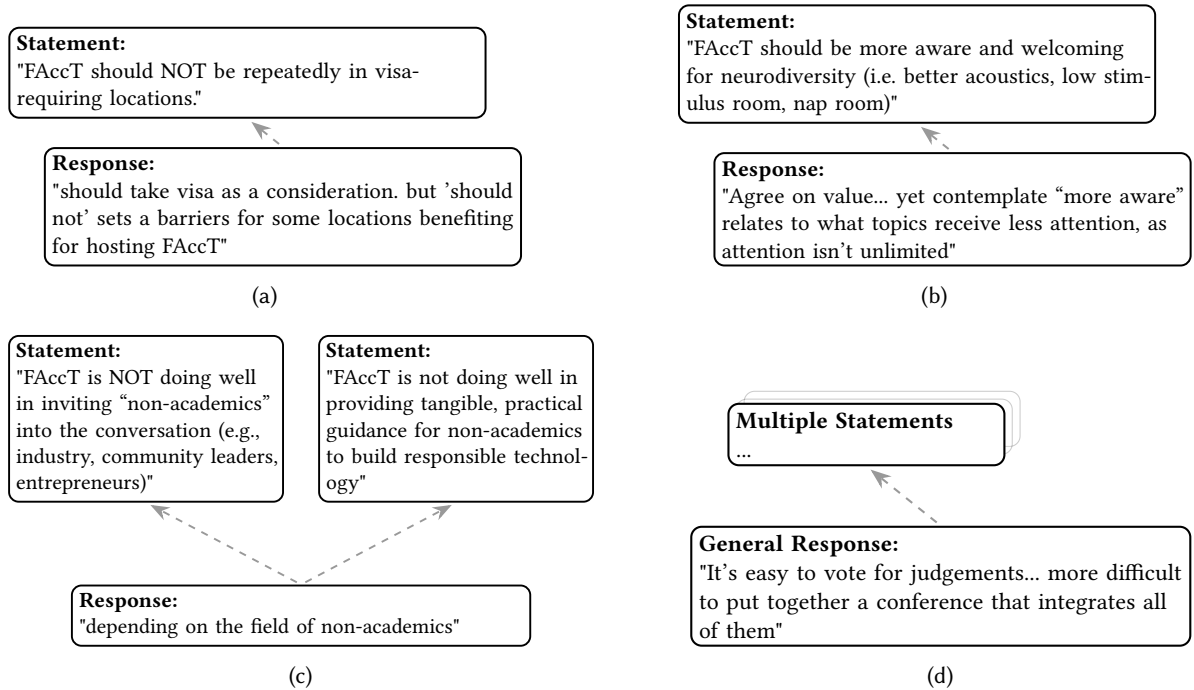
\begin{figure}[!h]
    \centering
    \tikzset{
        statement/.style={
            rectangle, draw=black, thick, rounded corners, 
            text width=6.5cm, align=left, font=\small, inner sep=3pt
        },
        smallstatement/.style={
            rectangle, draw=black, thick, rounded corners, 
            text width=3.8cm, align=left, font=\small, inner sep=3pt, fill=white
        },
        smallbox/.style={
            rectangle, draw=black, thick, rounded corners, 
            text width=3.5cm, align=left, font=\small, inner sep=3pt, fill=white
        },
        faded/.style={
            rectangle, draw=black, thin, rounded corners, 
            text width=3.5cm, inner sep=3pt, opacity=0.2, fill=white
        },
        arrow/.style={
            thick, dashed, draw=black!40, -{Stealth[scale=1.2]}, shorten >=2pt, shorten <=2pt
        }
    }

    \begin{subfigure}{0.44\textwidth}
        \centering
        \resizebox{\textwidth}{!}{
            \begin{tikzpicture}
                \node (v1) [statement] {
                    \textbf{Statement:}\\
                    "FAccT should NOT be repeatedly in visa-requiring locations."
                };
                \node (v2) [statement, below right=0.4cm and 0.5cm of v1.south west] {
                    \textbf{Response:}\\
                    "should take visa as a consideration. but 'should not' sets a barriers for some locations benefiting for hosting FAccT"
                };
                \draw [arrow] (v2.north) -- (v1.south);
            \end{tikzpicture}
        }
        \caption{}\label{fig:visas} 
    \end{subfigure}
    \hfill
    \begin{subfigure}{0.44\textwidth}
        \centering
        \resizebox{\textwidth}{!}{
            \begin{tikzpicture}
                \node (n1) [statement] {
                    \textbf{Statement:}\\
                    "FAccT should be more aware and welcoming for neurodiversity (i.e. better acoustics, low stimulus room, nap room)"
                }; 
                \node (n2) [statement, below right=0.4cm and 0.5cm of n1.south west] {
                    \textbf{Response:}\\
                    "Agree on value... yet contemplate “more aware” relates to what topics receive less attention, as attention isn’t unlimited"
                };
                \draw [arrow] (n2.north) -- (n1.south);
            \end{tikzpicture}
        }
        \caption{}\label{fig:neurodiversity} 
    \end{subfigure}

    \vspace{0.75em}

    \begin{subfigure}{0.5\textwidth}
        \centering
        \resizebox{\textwidth}{!}{
            \begin{tikzpicture}
                \node (s1) [smallstatement] {
                    \textbf{Statement:}\\
                    "FAccT is NOT doing well in inviting “non-academics” into the conversation (e.g., industry, community leaders, entrepreneurs)"
                };
                \node (s2) [smallstatement, right=0.5cm of s1] {
                    \textbf{Statement:}\\
                    "FAccT is not doing well in providing tangible, practical guidance for non-academics to build responsible technology"
                };
                
                \node (res) [statement, below=1.2cm of $(s1.south)!0.5!(s2.south)$] {
                    \textbf{Response:}\\
                    "depending on the field of non-academics"
                };

                \draw [arrow] (res.north) -- (s1.south);
                \draw [arrow] (res.north) -- (s2.south);
            \end{tikzpicture}
        }
        \caption{}\label{fig:ambiguous-target}
    \end{subfigure}
    \hfill
    \begin{subfigure}{0.43\textwidth} 
        \centering
        \resizebox{\textwidth}{!}{
            \begin{tikzpicture}
                \node (f2) [faded] at (6pt, 6pt) {\textbf{~}\\ \vspace{0.3em}...};
                \node (f1) [faded] at (3pt, 3pt) {\textbf{~}\\ \vspace{0.3em}...};
                \node (g1) [smallbox] at (0,0) {
                    \textbf{Multiple Statements}\\
                    ...
                };
                
                \node (g2) [statement, below=0.8cm of g1, xshift=1cm] {
                    \textbf{General Response:}\\
                    "It's easy to vote for judgements... more difficult to put together a conference that integrates all of them"
                };
                
                \draw [arrow] (g2.north) -- (g1.south);
            \end{tikzpicture}
        }
        \caption{}\label{fig:general-response} 
    \end{subfigure}

    \caption{\new{Successful and unsuccessful examples of active deliberation by participants: (a, b) direct responses, (c) response with an ambiguous target, and (d) general responses to multiple prior statements.}}
    \label{fig:statement-responses}
\end{figure}

\section{Discussion}
\new{Following \citet{bodker2022participatory}, we treat PD as more than just participation: PD is a process of mutual learning, participant influence on the problem space, and design-relevant consequence were organized in this case. \cite[Chs.\ 1--2]{bodker2022participatory}. On that baseline our case study, which witnessed iteration, negotiation and ultimately deliberation, is best understood as an \emph{agenda-setting} PD process, where participants co-framed seed statements through internal negotiation during the in-person CRAFT session, expanded the agenda through Polis and made patterns of agreement, disagreement and uncertainty visible through voting. Participants also used Polis for active deliberation, submitting new statements which responded to prior ones. Examples of this include statements responding to specific prior statements, e.g., ``FAccT should NOT be repeatedly in visa-requiring locations'' and ``should take visa as a consideration. but "should not" sets a barriers for some locations benefiting for hosting FAccT'', but also general responses, such as ``It's easy to vote for judgements about organizing a conference, more difficult to put together a conference that integrates all of them'' (cf.\ Fig.~\ref{fig:statement-responses}).

Our main design-centric outcome was the governance-facing report, which, following \citeauthor{bodker2022participatory}'s distinction between \emph{outputs}, \emph{outcomes} and \emph{impact} \cite[\S 9.4]{bodker2022participatory}, amounts to participatory outputs and plausible governance-relevant outcomes rather than full institutional impact. At the same time, our case study remained bounded in that the participants did not co-iterate the final report or share formal authority over implementation. The next cycle, accordingly, would involve further deliberation around contested mechanism and a visible leadership response loop. Such limits were not incidental but tied to the breadth-first, infrastructure-mediated form that made wider participation possible in the first place.}


\subsection{Scaling and infrastructuring in participatory design}
\label{subsec:scaling_infrastructuring}
\subsubsection{Typology of Scaling PD}
\new{To map the governance landscape of FAccT --- to make visible where tensions cluster, where alignments emerge and where disagreement runs deepest --- requires reaching beyond a single room of participants: indeed, a bounded workshop of a dozen people could not really surface the full terrain of concerns that exist across a distributed, international comunity. The PD work in our case study, therefore, depends fundamentaly on \emph{scale}, not as a background aspiration but as a constitutive feature of what makes the activity meaningful.}
\new{To understand what scaling means for participation, we turn to the typology developed by \citet{bossen_scaling_2025}, who distinguish between three partially overlapping dimensions along which scaling in PD unfolds}: socio-spatial, temporal and onto-epistemological \cite{bossen_scaling_2025}. For each dimension, we trace how the structured, asynchronous, infrastructure-mediated design created new openings for participation while simultaneously introducing specific constraint. Rather than focussing solely on whether the process had successfully scaled, we instead interrogate how the process shaped what became collectively legible and what opaque \emph{at scale}.
%

\textit{Socio-spatial.} Both our recruitment and participation took place via online platforms (the Slack channel and the Discus listserv for the former and Polis for the latter), expanding the potential reach to anyone in the FAccT community with internet access, not just those who had the funds and freedom of movement (visa). At the same time, participation still required internet access and proficiency in English, a barrier previously identified in other transnational large-scale PDs (e.g.\ \cite{queerinai2023queer}). Finally, we note that $ 10 $ participants voted on only one statement, suggesting that they drop immediately after starting participation. In sum, \emph{low-friction inclusion increases socio-spatial reach but not necessarily in a uniform manner.}

\textit{Temporal.} We observed both spikes in participation measured by both the cumulative numbers of statements and votes immediately after reminders, and the paucity of participants who have stayed engaged for long periods of time. Moreover, the vast majority of our participants joined after the conclusion of the in-person conference, and around half joined shortly after the announcement of the final deadline (cf.\ Fig.\ \ref{fig:participant_interval_full}). This suggests that sustained participation may be better measured by actively producing return, revisitation and maintenance facilitated through appropriate tools and infrastructures. At the same time, it required labour both on the part of the organizers to send out the reminders as well as of the community members to read the emails and messages to decide if they wanted to participate or continue participating. This implies that \emph{temporal scaling in large-scale PD requires well-engineered ``feedback loops'' with sufficient incentives.} \new{In addition, \emph{while sustained participation was challenged in the process, its asynchronous nature created low-friction inclusion that enhanced our reach}.}

\textit{Onto-epistemological.} An important effect of the anonymity \new{combined with sheer number of partipants which rendered it virtually impossible to recover the identity of the poster} has fostered safe spaces for controversial topics and often-neglected perspectives. The key lesson here is that \emph{onto-epistemological scaling can entail scaling epistemic safety: the conditions under which participants can surface controversial, dissenting, or historically neglected perspectives without being punished or humiliated for speaking their mind.} \new{This overcomes a limitation raised by Roland et al.~\cite{roland2017p}, who argued that current practices today over emphasize co-creation and agreeability over discovering tensions and conflicts among participants.} At the same time, this safety was purchased with the missingness of demographic information which could have enabled discovery of interesting patterns. On the other hand, the epistemic intelligibility of our PD was regulated by the requirement that the submitted statements should follow (``FAccT is [not] ...'', ``FAccT should [not] ...'', ``FAccT is [not] doing well in ...''). Therefore, \emph{large-scale PD can increase pluralism throughput by using lightweight generative constraints that invite multiple speech-acts (describing, prescribing, evaluating) without requiring consensus on definitions \emph{a priori}.}

Across these three dimensions, we also observe a platform-specific visibility regime: statements introduced earlier accrued more votes, suggesting that temporal scale is inseparable from ordering effects that shape what becomes collectively legible. In large-scale PD mediated by platforms, ``scale'' is therefore not only about reach, duration, or pluralism, but also about how participation is rendered visible and comparable through infrastructural choices (e.g., queueing, surfacing, and aggregation).

\new{

\subsubsection{Polis as a Platform}

Using Polis as a platform for PD comes with both limitations and opportunities. The main limitation of using Polis for PD is the character and depth of participation, where the platform's possible interactions offers only limited expression: statements are aimed to present simple ideas in a concise form, and response to statements is possible only through voting, or adding a new (standalone) statement. As a result there is a clear limit to the depth at which community mis/alignments can be identified or negotiated. Moreover, while we were able to observe successful deliberation via reflective statements in the participatory data (see Figs.~\ref{fig:visas} and \ref{fig:neurodiversity}), Polis does not directly support or encourage such reflection. This lack of support also lead to a failure mode we observed, where a participant seems to have thought they are directly responding to a statement, but it is unclear which statement they are responding to (see Fig.~\ref{fig:ambiguous-target}). On the other hand, where traditional approaches typically engage with only a handful of individuals who tend to be members of dominant groups~\cite{costanza2020design}, Polis has shown it scales not only in reach but also by supporting a plurality of voices. Moreover, under this setting, ideas are emerging from participants, offering a glimpse to tensions and agreements that are not always discussed or might be overly influenced in a top-down manner otherwise (e.g., by predefining a list of topics to discuss). Overall, as an information gathering tool for PD, its scaling affordances together with its limited interaction depth, Polis often plays a role during initial steps of the PD process in mapping the landscape of emerging needs -- informing on breadth, rather than depth, suitable to our context.

}

\subsubsection{FAccT as a Collective}
\label{sec:facct_collective}
In parallel, \citeauthor{huybrechts_collectives_2025} foreground \textit{collectives} and collective capabilities as central concerns in contemporary PD, showing how PD's engagements with complex issues such as mobility, migration or environmental justice rely on the formation and maintenance of diverse collectives that criss-cross institutional and spatial boundaries \cite{huybrechts_collectives_2025}. They define a collective as comprising of: (1) joint and distributed collective activities mediating understandings and solidarity; (2) a collectively articulated identity (made of internal and external practices); (3) a history and orientation towards future development or degradation; the involvement of more-than-human actors (both technological and ecological); and (4) the conscious articulation of implicated actors \cite[p.\ 54]{huybrechts_collectives_2025}.

These collectives are not pre-given ``communities'' waiting to be consulted but \new{constituted through ongoing practices and through the infrastructures that sustain them.}
Materially speaking, collectives are both enabled by and constitutive of \textit{infrastructuring} \cite{karasti2014infrastructuring,bodker_tying_2017}: meeting spaces, \new{listservs, Slack channels, and conference procedures} are continuously made and remade through the collective’s practices. Thus, infrastructuring involves cultivating transnational collectives that connect local spaces, activist networks, researchers, policy-makers and technological platforms, while recognising that these relations are fragile, contested and unevenly resourced. In the case of FAccT, a significant portion of activities that constitute the collective has taken place on the various associated online platforms outside of the strict confines of the annual conference. \new{In that sense, the label ``FAccT'' names not only the conference, but an extended sociotechnical formation of platforms, practices, and relationships. This perspective helps clarify what the present case can and cannot claim. The process documented here does not show that FAccT has already institutionalized this form of participation but rather identifies some of the conditions under which such institutioning may become possible, including organizer buy-in, participant ownership, follow-on deliberative stages, and a sustained mechanism for feeding the report back into conference governance \cite{huybrechts2017institutioning}. The main contribution is therefore not accomplished institutioning, but a clearer account of how a distributed conference collective might begin to build towards it.} 



\subsection{Alternative approaches}
To understand the contribution of the participatory process, we situate it within the context of traditional approaches typically used by venues such as FAccT to elicit feedback from community members. These approaches primarily include surveys and town hall meetings—methods common not only in academic conferences but also in workplace settings, and municipalities as will be discussed below. 

\subsubsection{Comparison with the post-conference survey}

There are several dimensions along which the contributions of the FAccT 2025 survey and the PD process described in this work differ. First, the primary addressees of the survey were attendees of the 2025 conference held in Greece, as the majority of questions focused on their recent conference experience. In contrast, the PD process was open to any FAccT community member. Second, the survey questions were authored by the organizers seeking feedback on specific concerns, whereas the statements in the PD process were authored by participants themselves. Third, the types of conversations differed substantially: 80\% of the 2025 survey questions were descriptive, focusing on the recent conference experience, with a smaller portion dedicated to normative and aspirational topics. In the PD process, only 39\% of the discussion was descriptive—addressing broad perceptions of FAccT—while 56\% focused on the vision participants aspired to for the community.\footnote{The complete list of survey questions and their assigned labels can be found in Table~\ref{tab:survey_questions} in the appendix, which underwent a similar labeling process as described for Table~\ref{tab:pd_statements_labeling}.} Finally, the survey had limited engagement with several themes that emerged prominently in the participatory design process, including the sensitivity to geographic \& cultural diversity, the care for environmental sustainability, the involvement of industry, and concerns about attendees' financial \& economic access.  
Based on this analysis the outcome of those two approaches is complementary.



\subsubsection{Comparison with the conference town hall}
At the FAccT 2025 town hall, conference attendees convened for an hour-long session that included both leadership updates and attendee feedback. While town halls are a valuable tool for eliciting attendee input, they come with inherent limitations, such as privileging the voices of those with public speaking skills and offering limited time for individuals to express their opinions. Like surveys, town halls may serve as a complementary means of gathering feedback.
One moment during the 2025 town hall was particularly memorable to the researchers (all of whom attended FAccT 2025): a community member raised concerns about FAccT's relationship with corporate funding. This statement was met with applause and a standing ovation, which may be interpreted as broad agreement with the voiced opinion. Through Polis, however, we learned that opinions on corporate funding are more complex than this moment might suggest. The community may benefit from examining this topic more deeply to articulate members' expectations regarding FAccT's relationship with industry and to identify where consensus can be achieved. In this sense, the town hall provided a glimpse into community sentiment, while the participatory design process enabled a more expansive and nuanced conversation on the topic.

\section{Limitations}

Our study has several limitations worth acknowledging. Similar to the other discussed approaches, there is a risk that sampling bias may adversely affect the representativeness of the results, given that participation was voluntary and open, which meant that people could, in theory, vote multiple times by using different physical devices, and also that not all participants voted on all statements, introducing a small risk of exposure affecting votes. With participation being anonymous, the degree of sampling bias is hard to assess. As the initial statements were created by the CRAFT-session attendees, they could have also overly influenced the discussion. Second, Polis as a platform offers a very specific (and limited) form of discourse, which has also been criticized in one of the participatory statements, stating ``Polis is not a useful way to have a conversation about these issues.'' This also highlights one of the strengths of participatory approaches, however, which is the option to gather feedback on and evolve the process itself. Nevertheless, Polis, as a platform, comes with various restrictions, including data that may be more challenging to aggregate than typical survey results. Given the grassroots source of our initiative, we were also not able to offer clear commitments on how participatory suggestions would be integrated into future conferences. Future participatory efforts--potentially designed in cooperation with conference organizers--could aim for more empowering forms of participation along \citeauthor{arnstein1969ladder}'s ``ladder of participation'' \cite{arnstein1969ladder}. 


\section{Conclusion}
In this work, we presented a case study of a reflexive large-scale PD --- the first of its kind at AI ethics/HCI conferences. We demonstrated both the utility and shortcomings of Polis, embedded in the wider FAccT online infrastructure, in surfacing value and novel insights on what the community thought about the current state and aspirational vision of FAccT. Finally, we suggested ways how our PD framework may be repeated and adapted to similar venues, demonstrating its potential of becoming part of the institution of conferences and other scholarly--activist collectives.


\clearpage

\section*{Acknowledgements}

We would like to express our heartfelt thanks to all the participants who generously shared their views via polis and especially everyone who attended the CRAFT Session. Your contributions were essential to the success of this research, and we deeply appreciate the time and effort you contributed to this project.

This work is supported by the DAAD programme Konrad Zuse Schools of Excellence in Artificial Intelligence, sponsored by the Federal Ministry of Education and Research, the Munich Center for Machine Learning and the seed funding program at the Mannheim Center for European Social Research (MZES) at the University of Mannheim. Computational resources were provided by Google TPU Research Cloud (TRC).

\section*{Generative AI statement}
As explained in \S~\ref{sec:topic}, we used LLMs to assist with the thematic analysis. We also used LLM to proofread our texts and to assist in the development of analysis code \new{and figures}. 

\bibliographystyle{ACM-Reference-Format}
\bibliography{refs.bib}

@incollection{bodker2022participatory,
  title={What is participatory design?},
  author={B{\o}dker, Susanne and Dindler, Christian and Iversen, Ole S and Smith, Rachel C},
  booktitle={Participatory design},
  pages={5--13},
  year={2022},
  publisher={Springer}
}

@inproceedings{karasti2014infrastructuring,
  title={Infrastructuring in participatory design},
  author={Karasti, Helena},
  booktitle={Proceedings of the 13th Participatory Design Conference: Research Papers-Volume 1},
  pages={141--150},
  year={2014}
}

@article{huybrechts2017institutioning,
  title={Institutioning: Participatory design, co-design and the public realm},
  author={Huybrechts, Liesbeth and Benesch, Henric and Geib, Jon},
  journal={CoDesign},
  volume={13},
  number={3},
  pages={148--159},
  year={2017},
  publisher={Taylor \& Francis}
}

@inproceedings{tseng2025ownership,
  title={" Ownership, Not Just Happy Talk": Co-Designing a Participatory Large Language Model for Journalism},
  author={Tseng, Emily and Young, Meg and Le Qu{\'e}r{\'e}, Marianne Aubin and Rinehart, Aimee and Suresh, Harini},
  booktitle={Proceedings of the 2025 ACM Conference on Fairness, Accountability, and Transparency},
  pages={3119--3130},
  year={2025}
}

@article{young2024participation,
  title={Participation versus scale: Tensions in the practical demands on participatory AI},
  author={Young, Meg and Ehsan, Upol and Singh, Ranjit and Tafesse, Emnet and Gilman, Michele and Harrington, Christina and Metcalf, Jacob},
  journal={First Monday},
  year={2024}
}

@inproceedings{queerinai2023queer,
  title={Queer in AI: A case study in community-led participatory AI},
  author={Queerinai, Organizers Of and Ovalle, Anaelia and Subramonian, Arjun and Singh, Ashwin and Voelcker, Claas and Sutherland, Danica J and Locatelli, Davide and Breznik, Eva and Klubicka, Filip and Yuan, Hang and others},
  booktitle={Proceedings of the 2023 ACM Conference on Fairness, Accountability, and Transparency},
  pages={1882--1895},
  year={2023}
}

@inproceedings{septiandri2023weird,
  title={WEIRD FAccTs: How western, educated, industrialized, rich, and democratic is FAccT?},
  author={Septiandri, Ali Akbar and Constantinides, Marios and Tahaei, Mohammad and Quercia, Daniele},
  booktitle={Proceedings of the 2023 ACM Conference on Fairness, Accountability, and Transparency},
  pages={160--171},
  year={2023}
}

@inproceedings{schroeder2025disclosure,
  title={Disclosure without Engagement: An Empirical Review of Positionality Statements at FAccT},
  author={Schroeder, Hope and Pareek, Akshansh and Barocas, Solon},
  booktitle={Proceedings of the 2025 ACM Conference on Fairness, Accountability, and Transparency},
  pages={1195--1210},
  year={2025}
}

@inproceedings{laufer2022four,
  title={Four years of FAccT: A reflexive, mixed-methods analysis of research contributions, shortcomings, and future prospects},
  author={Laufer, Benjamin and Jain, Sameer and Cooper, A Feder and Kleinberg, Jon and Heidari, Hoda},
  booktitle={Proceedings of the 2022 ACM conference on fairness, accountability, and transparency},
  pages={401--426},
  year={2022}
}

@inproceedings{sloane2022participation,
  title={Participation is not a design fix for machine learning},
  author={Sloane, Mona and Moss, Emanuel and Awomolo, Olaitan and Forlano, Laura},
  booktitle={Proceedings of the 2nd ACM Conference on Equity and Access in Algorithms, Mechanisms, and Optimization},
  pages={1--6},
  year={2022}
}

@inproceedings{birhane2022power,
  title={Power to the people? Opportunities and challenges for participatory AI},
  author={Birhane, Abeba and Isaac, William and Prabhakaran, Vinodkumar and Diaz, Mark and Elish, Madeleine Clare and Gabriel, Iason and Mohamed, Shakir},
  booktitle={Proceedings of the 2nd ACM Conference on Equity and Access in Algorithms, Mechanisms, and Optimization},
  pages={1--8},
  year={2022}
}

@inproceedings{huang2024collective,
  title={Collective constitutional ai: Aligning a language model with public input},
  author={Huang, Saffron and Siddarth, Divya and Lovitt, Liane and Liao, Thomas I and Durmus, Esin and Tamkin, Alex and Ganguli, Deep},
  booktitle={Proceedings of the 2024 ACM Conference on Fairness, Accountability, and Transparency},
  pages={1395--1417},
  year={2024}
}

@inproceedings{delgado2023participatory,
  title={The participatory turn in ai design: Theoretical foundations and the current state of practice},
  author={Delgado, Fernando and Yang, Stephen and Madaio, Michael and Yang, Qian},
  booktitle={Proceedings of the 3rd ACM Conference on Equity and Access in Algorithms, Mechanisms, and Optimization},
  pages={1--23},
  year={2023}
}

@incollection{bossen_scaling_2025,
  author    = {Claus Bossen and Reem Talhouk and John Vines},
  title     = {Scaling Participatory Design},
  booktitle = {Routledge International Handbook of Contemporary Participatory Design},
  editor    = {Rachel Charlotte Smith and Daria Loi and Heike Winschiers-Theophilus and Liesbeth Huybrechts and Jesper Simonsen},
  publisher = {Routledge},
  address   = {London},
  year      = {2025},
  doi       = {10.4324/9781003334330-3},
}

@incollection{huybrechts_collectives_2025,
  author    = {Liesbeth Huybrechts and Maurizio Teli and Joanna Saad{-}Sulonen and Mariacristina Sciannamblo},
  title     = {Collectives and Participatory Design},
  booktitle = {Routledge International Handbook of Contemporary Participatory Design},
  editor    = {Rachel Charlotte Smith and Daria Loi and Heike Winschiers-Theophilus and Liesbeth Huybrechts and Jesper Simonsen},
  publisher = {Routledge},
  address   = {London},
  year      = {2025},
  doi       = {10.4324/9781003334330-4},
}

@inproceedings{zahlsen_challenges_2022,
  author    = {{\O}ivind Klungseth Zahlsen and Elena Parmiggiani and Yngve Dahl},
  title     = {Challenges of Scaling Participatory Design: A Systematic Literature Review},
  booktitle = {Proceedings of the 34th Australian Conference on Human-Computer Interaction (OzCHI '22)},
  publisher = {ACM},
  address   = {New York, NY},
  year      = {2022},
  doi       = {10.1145/3572921.3572924},
}

@article{bodker_tying_2017,
  author    = {Susanne B{\o}dker and Christian Dindler and Ole Sejer Iversen},
  title     = {Tying Knots: Participatory Infrastructuring at Work},
  journal   = {Computer Supported Cooperative Work (CSCW)},
  year      = {2017},
  volume    = {26},
  number    = {1--2},
  pages     = {245--273},
  doi       = {10.1007/s10606-017-9268-y},
}

@article{wacnik_participatory_2025,
  author    = {Peter Wacnik and Shanna R. Daly and Aditi Verma},
  title     = {Participatory Design: A Systematic Review and Insights for Future Practice},
  journal   = {Design Science},
  year      = {2025},
  volume    = {11},
  pages     = {e21},
  doi       = {10.1017/dsj.2025.10009},
}

@article{small2023opportunities,
  title={Opportunities and risks of LLMs for scalable deliberation with Polis},
  author={Small, Christopher T and Vendrov, Ivan and Durmus, Esin and Homaei, Hadjar and Barry, Elizabeth and Cornebise, Julien and Suzman, Ted and Ganguli, Deep and Megill, Colin},
  journal={arXiv preprint arXiv:2306.11932},
  year={2023}
}

@article{small2021polis,
  title={Polis: Scaling deliberation by mapping high dimensional opinion spaces},
  author={Small, Christopher and Bjorkegren, Michael and Erkkil{\"a}, Timo and Shaw, Lynette and Megill, Colin},
  journal={Recerca: revista de pensament i an{\`a}lisi},
  volume={26},
  number={2},
  year={2021},
  publisher={Universitat Jaume I Servei de Comunicacio i Publicacions}
}

@article{arnstein1969ladder,
  title = {A {{Ladder Of Citizen Participation}}},
  author = {Arnstein, Sherry R.},
  year = 1969,
  month = jul,
  journal = {Journal of the American Institute of Planners},
  volume = {35},
  number = {4},
  pages = {216--224},
  publisher = {Routledge},
  issn = {0002-8991},
  doi = {10.1080/01944366908977225},
  url = {https://doi.org/10.1080/01944366908977225},
  urldate = {2025-02-25}
}

@book{costanza2020design,
  title={Design justice: Community-led practices to build the worlds we need},
  author={Costanza-Chock, Sasha},
  year={2020},
  publisher={MIT press}
}

@article{roland2017p,
  title={P for Platform. Architectures of large-scale participatory design},
  author={Roland, Lars Kristian and Sanner, Terje Aksel and S{\ae}b{\o}, Johan Ivar and Monteiro, Eric},
  journal={Scandinavian Journal of Information Systems},
  volume={29},
  number={2},
  pages={1},
  year={2017}
}

\appendix
\clearpage

\section{PD Statement classification table}

\begin{longtable}{@{}cp{5.5cm}lp{4.5cm}@{}}
\caption{Polis Statements with Classification and Theme Labels}
\label{tab:pd_statements_labeling} \\
    \toprule
    \textbf{ID} & \textbf{Statement} & \textbf{Conversation Focus} & \textbf{Theme(s)} \\
    \midrule
    \endfirsthead
    
    \multicolumn{4}{c}{\tablename\ \thetable{} -- continued from previous page} \\
    \toprule
    \textbf{ID} & \textbf{Statement} & \textbf{Type} & \textbf{Theme(s)} \\
    \midrule
    \endhead
    
    \midrule
    \multicolumn{4}{r}{Continued on next page} \\
    \endfoot
    
    \bottomrule
    \endlastfoot
    
    0 & FAccT should be open and welcoming community & Normative & Community \& Values; Inclusivity \& Accessibility \\
    1 & FAccT should be a generator for alternative solidarity structures (such as mutual aid) & Normative & Community \& Values; Practical Impact \\
    2 & FAccT is NOT doing well in inviting ``non-academics'' into the conversation (e.g., industry, community leaders, entrepreneurs) & Descriptive & Inclusivity \& Accessibility; Industry Engagement \\
    3 & FAccT should have poster sessions & Normative & Conference Format \& Programming \\
    4 & FAccT is not inclusive to the local country/community & Descriptive & Geographic \& Cultural Diversity; Inclusivity \& Accessibility \\
    5 & FAccT is not doing well in providing tangible, practical guidance for non-academics to build responsible technology & Descriptive & Practical Impact \\
    6 & FAccT should have more opportunities to meet people (i.e.\ speed dating, poster sessions) & Normative & Conference Format \& Programming \\
    7 & FAccT is doing well in implementing requests from previous years into future conferences & Descriptive & Governance \& Operations \\
    8 & FAccT should invite more industry (as visitors) but not give them a platform & Normative & Industry Engagement \\
    9 & FAccT should be free for PhD students & Normative & Financial \& Economic Access \\
    10 & FAccT should have different prices for different countries' income group & Normative & Financial \& Economic Access; Geographic \& Cultural Diversity; Inclusivity \& Accessibility \\
    11 & FAccT is building a community that shares similar values & Descriptive & Community \& Values \\
    12 & FAccT is not advertised locally (even among academics) & Descriptive & Geographic \& Cultural Diversity \\
    13 & FAccT is mostly focused on a single discipline (CS) & Descriptive & Disciplinary Scope \\
    14 & FAccT is the most multidisciplinary conference in Computer Science & Descriptive & Disciplinary Scope \\
    15 & FAccT should be more aware and welcoming for neurodiversity (i.e.\ better acoustics in the building, low stimulus room, nap room) & Normative & Inclusivity \& Accessibility \\
    16 & FAccT is not inclusive to non-academics & Descriptive & Disciplinary Scope; Inclusivity \& Accessibility \\
    17 & FAccT should be vegetarian & Normative & Environmental Sustainability \\
    18 & FAccT is doing well in respectful exchange of interdisciplinary topics & Descriptive & Disciplinary Scope \\
    19 & FAccT should NOT be exclusionary & Normative & Inclusivity \& Accessibility \\
    20 & FAccT should NOT be sponsored by big corporates & Normative & Industry Engagement \\
    21 & FAccT is not doing well in creating opportunities for participatory design, prototyping, and building/creating together AT the conference & Descriptive & Conference Format \& Programming; Practical Impact \\
    22 & FAccT is doing well in sticking to their values & Descriptive & Community \& Values \\
    23 & FAccT is doing well at inclusivity & Descriptive & Community \& Values; Inclusivity \& Accessibility \\
    24 & FAccT is doing well in selecting diverse locations that are fun and culturally enriched & Descriptive & Geographic \& Cultural Diversity \\
    25 & FAccT should NOT be centred around western narratives (e.g.\ US right now) and political atmosphere & Normative & Geographic \& Cultural Diversity \\
    26 & FAccT should NOT contribute to plastic waste & Normative & Environmental Sustainability \\
    27 & FAccT is a useful space to meet future collaborators & Descriptive & Community \& Values; Conference Format \& Programming \\
    28 & FAccT should NOT dismiss environmental footprint of research & Normative & Environmental Sustainability \\
    29 & FAccT is doing well in providing different formats and ways to connect like socials, CRAFTs, etc. & Descriptive & Conference Format \& Programming \\
    30 & FAccT should NOT be repeatedly in visa-requiring locations & Normative & Geographic \& Cultural Diversity; Inclusivity \& Accessibility \\
    31 & FAccT should be multilingual & Normative & Geographic \& Cultural Diversity; Inclusivity \& Accessibility \\
    32 & FAccT should diversify its epistemology & Normative & Disciplinary Scope \\
    33 & FAccT is US centric in content and attendees & Descriptive & Geographic \& Cultural Diversity \\
    34 & FAccT should engage in real participatory design & Normative & Conference Format \& Programming; Practical Impact \\
    35 & FAccT should have less simultaneous paper tracks (offering the choice between papers, CRAFT and tutorial) & Normative & Conference Format \& Programming \\
    36 & FAccT should remain volunteer-run (organizers are uncompensated) & Normative & Governance \& Operations \\
    37 & FAccT should take industry funding, but only the minimum required to operate & Normative & Financial \& Economic Access; Industry Engagement \\
    38 & FAccT should reduce sponsorship reliance to a minimum by no longer providing financial support to any attendees & Normative & Financial \& Economic Access; Industry Engagement \\
    39 & FAccT should have an ``interventions/visioning'' track/workshop that makes space for new proposals and a critical discussion of them & Normative & Conference Format \& Programming \\
    40 & Because there aren't other conferences with this focus, FAccT must remain a top venue for research in computing and social justice & Normative & Community \& Values; Disciplinary Scope \\
    41 & Academics often need publications to justify attendance. FAccT should add publications (e.g.\ workshops) to broaden audience esp.\ students & Normative & Conference Format \& Programming; Financial \& Economic Access \\
    42 & FAccT should support more cross-disciplinary meetings at the conference, not just in the socials & Normative & Conference Format \& Programming; Disciplinary Scope \\
    43 & Regarding whether FAccT remains volunteer-run, I think we shall also think about ``what is the role of compensation'', the necessity & Normative & Governance \& Operations \\
    44 & Agree on value of neurodiversity, yet, contemplate ``more aware'' relates to what topics receive less attention, as attention isn't unlimited & Other & Inclusivity \& Accessibility; Meta-Feedback \\
    46 & Better question is the alternatives (how to) not center around western narratives & Normative & Geographic \& Cultural Diversity \\
    48 & Should take visa as a consideration, but ``should not'' sets barriers for some locations benefiting from hosting FAccT & Normative & Geographic \& Cultural Diversity; Inclusivity \& Accessibility \\
    49 & Polis is not a useful way to have a conversation about these issues & Other & Meta-Feedback \\
    50 & Many of the ideas in here are things that FAccT already does & Descriptive & Meta-Feedback \\
    51 & Financial support to attendees is a paramount expression of our values & Descriptive & Community \& Values; Financial \& Economic Access \\
    52 & It's easy to vote for judgements about organizing a conference, more difficult to put together a conference that integrates all of them & Other & Governance \& Operations; Meta-Feedback \\
    53 & FAccT should be more inclusive of people \& topics important outside the US, Europe \& scrutinize prominence of corporate papers \& narratives & Normative & Geographic \& Cultural Diversity; Inclusivity \& Accessibility; Industry Engagement \\
    54 & Poster sessions provide value as an easy way to randomly talk to anyone and start a conversation, sometimes difficult otherwise at FAccT & Descriptive & Conference Format \& Programming \\
    55 & Three times I've been to FAccT ('18, '23, '24) tech industry influence (e.g.\ who runs workshops) has been noticeable \& sometimes problematic & Descriptive & Industry Engagement \\
    56 & FAccT could require panels \& workshops to be designed/led by representatives of that year's region. '24 had decent regional representation & Normative & Conference Format \& Programming; Geographic \& Cultural Diversity \\
    57 & In 3 years I've been to FAccT it had plenty of industry input through the papers chosen \& who ran or spoke at panels \& workshops & Descriptive & Conference Format \& Programming; Industry Engagement \\
    58 & FAccT should also run local meetups (EU/UK/Asia-Pacific) in case attendees fail to attend due to visa reasons & Normative & Conference Format \& Programming; Geographic \& Cultural Diversity; Inclusivity \& Accessibility \\
    59 & FAccT should not get money from big corporations & Normative & Industry Engagement \\
    60 & A FAccT paper submission is more likely to be rejected if it takes a critical perspective on past published work & Descriptive & Conference Format \& Programming; Disciplinary Scope \\
\end{longtable}

section{Survey question classification}

\begin{table}[htbp]
\centering
\caption{FAccT Survey 2025 Questions with Classification and Theme Labels}
\label{tab:survey_questions}
\scriptsize
\begin{tabular}{@{}cp{6cm}lp{4.5cm}@{}}
    \toprule
    \textbf{ID} & \textbf{Survey Question} & \textbf{Conversation Focus} & \textbf{Theme(s)} \\
    \midrule
    1 & I attended presentations from scholarly disciplines (and areas of practice) other than my own & Descriptive & Disciplinary Scope \\
    2 & I had productive exchanges with scholars and practitioners from other disciplines than my own & Descriptive & Disciplinary Scope \\
    3 & Please rate your conference experience & Descriptive & Conference Format \& Programming \\
    4 & How would you rate the keynotes, community keynote, and keynote panel? & Descriptive & Conference Format \& Programming \\
    5 & How would you rate the CRAFT program? & Descriptive & Conference Format \& Programming \\
    6 & How would you rate the Tutorials program? & Descriptive & Conference Format \& Programming \\
    7 & How would you rate the Proceedings track (the papers) & Descriptive & Conference Format \& Programming \\
    8 & How would you rate the Socials? & Descriptive & Community \& Values; Conference Format \& Programming \\
    9 & Was the Town Hall effective? & Descriptive & Conference Format \& Programming; Governance \& Operations \\
    10 & How could we improve any of these aspects of the program? & Normative & Conference Format \& Programming; Governance \& Operations \\
    11 & I felt valued and respected at the conference & Descriptive & Community \& Values; Inclusivity \& Accessibility \\
    12 & I felt safe participating and sharing my perspectives & Descriptive & Community \& Values; Inclusivity \& Accessibility \\
    13 & I think the tone of moderators, speakers, panelists, attendees and organisers created an inclusive atmosphere & Descriptive & Community \& Values; Inclusivity \& Accessibility \\
    14 & I was able to learn from perspectives that I do not usually encounter in my other professional contexts & Descriptive & Disciplinary Scope; Inclusivity \& Accessibility \\
    15 & The organisers featured a mix of people from across industry, civil society and academia, including those directly affected by algorithmic harms & Descriptive & Disciplinary Scope; Inclusivity \& Accessibility; Industry Engagement \\
    16 & Did you attend the conference in-person or online? & Descriptive & Conference Format \& Programming; Inclusivity \& Accessibility \\
    17 & I liked the venue (the Athens Conservatory) & Descriptive & Conference Format \& Programming; Geographic \& Cultural Diversity \\
    18 & I liked the location (Athens, Greece) & Descriptive & Geographic \& Cultural Diversity \\
    19 & I liked the catering (lunches and coffee breaks) & Descriptive & Conference Format \& Programming \\
    20 & I felt supported finding help if I didn't know where something was or how to do something & Descriptive & Governance \& Operations; Inclusivity \& Accessibility \\
    21 & I had enough time for informal social interaction & Descriptive & Community \& Values; Conference Format \& Programming \\
    22 & In general, how do you think we could improve in-person conference delivery? & Normative & Conference Format \& Programming \\
    23 & Tell us about some highlights of the conference experience for you. What worked & Descriptive & Other \\
    24 & Next please tell us what we could have done better, or anything you'd like to see change & Normative & Other \\
    25 & Do you have any broader observations you'd like to share about FAccT (its governance and policies, how it is funded, or other matters) & Both & Governance \& Operations; Industry Engagement \\
    26 & Is there anything else you'd like to share with the organising committee? & Both & Meta-Feedback \\
\bottomrule
\end{tabular}
\end{table}

\section{Participatory Materials}
\label{sec:materials}

\textit{Each participant group received two pieces of paper with the following information and two matched, opposing prefixes e.g. "FAccT is...", "FAccT is not...".}

Please write your seed statements below for discussion. Collect a list of statements first, review them and then submit them to the Google form once you are happy with your selection.

What makes a good statement?

\begin{enumerate}
    \item Short, 1-2 sentences.
    \item Each comment has to be on a single point.
    \item Use simple, clear language.
    \item Make sure there's a diversity of opinions represented — as many different responses to the prompt as possible.
    \item Include both controversial and potentially consensus statements
\end{enumerate}

\textit{[List of 10 empty lines, each starting with the given prefix e.g. "FAccT is..."]}

\section{Theme-based analysis voting patterns}
\label{sec:vote_theme}

\paragraph{Community \& Values} Participants expressed strong consensus with high vote count that FAccT should be an open and welcoming community (89\% agree). With similarly high participation rate, a majority agreed that FAccT serves as a useful space to meet future collaborators (72\% agree), is building a community with shared values (58\% agree), and is doing well at sticking to those values (51\% agree, though with a notable 40\% pass rate suggesting some uncertainty). However, opinions were mixed on whether FAccT is doing well at inclusivity (47\% agree, 37\% pass) and whether the conference should serve as a generator for alternative solidarity structures such as mutual aid (44\% agree, 20\% disagree)—both with high participation rate, indicating these are salient but unresolved questions. Among statements with fewer voters, there was majority consensus that FAccT occupies a unique and necessary role as a top venue for computing and social justice research (72\% agree, moderate vote count) and that financial support to attendees is an important expression of community values (69\% agree, low vote count). Taken together, these patterns suggest a community firmly united around ideals of openness and collaboration, with broad but less certain agreement on how well FAccT lives up to its values, and genuine division on the scope of its social mission. 

\paragraph{Conference Format \& Programming} Participants showed strong agreement with high participation rate that FAccT is a useful space to meet future collaborators (72\% agree) and should offer more opportunities to connect, such as speed dating and poster sessions (68\% agree). There was also majority consensus that FAccT is doing well in providing varied formats like socials and CRAFTs (63\% agree) and that poster sessions specifically should be part of the program (66\% agree). Among those who voted on lower-vote count statements, there was strong consensus that poster sessions provide particular value as an easy way to start conversations (76\% agree, low vote count), and majority support for ideas like local regional meetups for those facing visa barriers (71\% agree), more cross-disciplinary meetings beyond socials (65\% agree), and additional publication opportunities such as workshops to help academics justify attendance (58\% agree). However, opinions were mixed on several programming questions: whether FAccT should engage in real participatory design (48\% agree, 40\% pass), reduce simultaneous paper tracks (40\% agree, 21\% disagree), or require panels and workshops to be led by representatives from that year's host region (42\% agree, 39\% pass). There was also high uncertainty about one participant's comment on whether FAccT has had plenty of industry input in recent years (37\% agree, 63\% pass), and mixed views on whether the conference is falling short in creating opportunities for participatory design and building together on-site (38\% agree, 44\% pass). Overall, participants broadly value connection and collaboration opportunities at FAccT, while remaining divided or uncertain about structural changes to programming and the role of participatory and regional approaches.

\paragraph{Disciplinary Scope} Participants showed majority consensus with high participation rate that FAccT is doing well in fostering respectful exchange across interdisciplinary topics (67\% agree). With moderate vote count, there was also majority agreement that FAccT occupies a unique position and must remain a top venue for computing and social justice research (72\% agree), and that the conference should support more cross-disciplinary meetings beyond just socials (65\% agree). However, opinions were more divided on questions of disciplinary identity. The claim that FAccT is mostly focused on a single discipline (CS) was rejected by a plurality (47\% disagree, 25\% agree), though with high participation rate, suggesting participants broadly see FAccT as more than a CS conference. At the same time, the proposition that FAccT is the most multidisciplinary conference in computer science drew mixed responses (47\% agree, 35\% pass), indicating some ambivalence about how strongly to make that claim. There was also considerable uncertainty around whether FAccT is inclusive to non-academics (29\% agree, 24\% disagree, 47\% pass) and whether it should diversify its epistemology (47\% agree, 45\% pass)—both high-vote count statements where many participants chose to pass rather than take a firm stance. Overall, the community appears confident that FAccT handles interdisciplinary exchange respectfully and holds a distinctive place in the field, but remains uncertain or divided on the depth of its disciplinary diversity and openness to non-academic perspectives.

\paragraph{Environmental Sustainability} Participants expressed majority consensus with high participation rate on two environmental sustainability concerns: that FAccT should not dismiss the environmental footprint of research (68\% agree) and that the conference should not contribute to plastic waste (64\% agree). However, the more prescriptive proposal that FAccT should be vegetarian drew mixed opinions (32\% agree, 46\% disagree), suggesting that while the community broadly supports environmental awareness in principle, there is resistance to specific mandates around food choices. The voting patterns indicate a community that values sustainability as a consideration but is divided on how far such commitments should extend into individual or collective practices at the conference.

\paragraph{Financial \& Economic Access} Participants showed majority consensus with high participation rate that FAccT should have tiered pricing based on countries' income levels (68\% agree), reflecting support for economic accessibility across global contexts. There was also strong majority disagreement with the proposal to reduce sponsorship reliance by eliminating financial support to attendees (72\% disagree, moderate vote count), reinforcing that attendee support is valued. Indeed, among those who voted on a lower-vote count statement, there was majority consensus that financial support to attendees is a paramount expression of the community's values (69\% agree). With moderate vote count, participants also supported adding publication opportunities such as workshops to help academics—especially students—justify attendance (58\% agree). However, opinions were mixed on whether FAccT should be free for PhD students (38\% agree, 36\% disagree) and whether the conference should accept industry funding only at the minimum level required to operate (38\% agree, 28\% disagree, 35\% pass). Overall, the community clearly prioritizes financial accessibility and attendee support, but remains divided on the specific mechanisms—such as free student registration or limiting industry funding—for achieving these goals.

\paragraph{Geographic \& Cultural Diversity} Participants showed majority consensus with high participation rate that FAccT is currently US-centric in content and attendees (59\% agree) and that the conference should not be centred around Western narratives and political atmosphere (53\% agree). At the same time, there was majority agreement that FAccT is doing well in selecting diverse, culturally enriched locations (58\% agree) and should offer tiered pricing based on countries' income levels (68\% agree). With moderate vote count, participants supported considering visa requirements when selecting locations while acknowledging that strict restrictions could disadvantage some regions from hosting (54\% agree), and with low vote count there was strong majority support for FAccT being more inclusive of people and topics important outside the US and Europe (72\% agree) as well as running regional meetups for those facing visa barriers (71\% agree). However, several proposals drew mixed opinions: whether FAccT should avoid repeatedly holding the conference in visa-requiring locations (47\% agree, 38\% pass), whether it should be multilingual (44\% agree, 26\% disagree), and whether panels and workshops should be required to be led by representatives from that year's host region (42\% agree, 39\% pass). There was also high uncertainty—despite high participation rate—about whether FAccT is inclusive to the local country or community (55\% pass) and whether the conference is adequately advertised locally even among academics (55\% pass), suggesting many participants felt unable to evaluate these claims. Overall, the community recognizes a US and Western tilt in FAccT's current orientation and broadly supports greater geographic and cultural inclusivity, but remains divided or uncertain on specific mechanisms for achieving this shift.

\paragraph{Governance \& Operations} Participants demonstrated high uncertainty with high participation rate about whether FAccT is doing well in implementing requests from previous years into future conferences (37\% agree, 57\% pass), suggesting many voters felt unable to assess the conference's responsiveness to community feedback. Opinions were mixed on whether FAccT should remain volunteer-run with uncompensated organizers (45\% agree, 21\% disagree), and there was similarly high uncertainty at moderate participation rate about how to think through the role of compensation in this context (36\% agree, 54\% pass). However, among those who voted on a lower-participation rate statement, there was strong consensus acknowledging that while it is easy to vote on judgments about organizing a conference, it is far more difficult to actually put together a conference that integrates all of them (79\% agree). Overall, the voting patterns reveal a community that is largely uncertain or divided on governance questions, while recognizing the practical complexity of conference organization.

\paragraph{Inclusivity \& Accessibility} Participants expressed strong consensus with high participation rate that FAccT should be an open and welcoming community (89\% agree) and majority consensus that it should not be exclusionary (68\% agree). There was also broad support for concrete accessibility measures: tiered pricing based on countries' income levels (68\% agree, high participation rate), greater awareness and accommodation for neurodiversity such as better acoustics and low-stimulus rooms (56\% agree, high participation rate), and consideration of visa requirements when selecting locations while recognizing that overly strict rules could prevent some regions from hosting (54\% agree, moderate participation rate). Among lower-participation rate statements, there was strong majority support for FAccT being more inclusive of people and topics outside the US and Europe (72\% agree) and for running regional meetups to accommodate those facing visa barriers (71\% agree). However, opinions were mixed on several questions about FAccT's current performance and future direction: whether the conference is doing well at inclusivity overall (47\% agree, 37\% pass), whether it is inclusive to non-academics (29\% agree, 47\% pass), whether it is successfully inviting non-academics into the conversation (40\% agree, 27\% disagree), whether it should avoid repeatedly holding the conference in visa-requiring locations (47\% agree, 38\% pass), and whether it should be multilingual (44\% agree, 26\% disagree). There was also high uncertainty about whether FAccT is inclusive to local countries and communities (55\% pass) and about how to balance attention to neurodiversity against limited capacity for new topics (62\% pass). Overall, the community is united around aspirational values of openness and non-exclusion, and supports several practical accessibility measures, but remains divided or uncertain on how well FAccT currently delivers on inclusivity and on some of the specific mechanisms for broadening access.

\paragraph{Industry Engagement} The community showed clear division on questions of industry engagement, with nearly all high-participation rate statements yielding mixed opinions. Participants were split on whether FAccT should refuse sponsorship from big corporations (34\% agree, 43\% disagree), whether it should accept industry funding only at the minimum level required to operate (38\% agree, 28\% disagree), and whether it should invite more industry visitors without giving them a platform (43\% agree, 40\% pass). There was also no consensus on whether FAccT is currently doing well at inviting non-academics—including industry, community leaders, and entrepreneurs—into the conversation (40\% agree, 27\% disagree). However, participants did show strong majority disagreement with the proposal to reduce sponsorship reliance by eliminating financial support to attendees (72\% disagree, moderate participation rate), underscoring that attendee support remains a priority even amid debates about corporate involvement. Among lower-participation rate statements, there was majority consensus that FAccT should scrutinize the prominence of corporate papers and narratives (72\% agree), but high uncertainty about individual observations regarding industry influence: one claim that tech industry influence has been noticeable and sometimes problematic drew 50\% pass, while another asserting that FAccT has had plenty of industry input drew 63\% pass, suggesting many voters lacked the experience or information to evaluate these observations. Overall, the community is actively grappling with the role of industry at FAccT—resistant to eliminating attendee support but without consensus on how much corporate funding or presence is appropriate.

\paragraph{Meta-Feedback} Participants offered reflections on the feedback process itself, revealing a mix of uncertainty and self-awareness. There was strong consensus among those who voted that it is easy to vote on judgments about organizing a conference but far more difficult to actually put together a conference that integrates all of them (79\% agree, low participation rate)—an acknowledgment of the gap between expressing preferences and implementing them. However, high uncertainty characterized other meta-level observations: whether the Polis platform is a useful way to have these conversations (40\% agree, 51\% pass), whether many of the ideas raised are things FAccT already does (30\% agree, 56\% pass), and whether calls for greater awareness on topics like neurodiversity need to be weighed against the reality that attention is limited (25\% agree, 62\% pass). The prevalence of pass votes across these statements suggests participants were hesitant to render firm judgments on the process and its relationship to existing practices, which may align with the acknowledged difficulty of translating community input into conference design.

\paragraph{Practical Impact} Participants showed mixed opinions across all statements related to practical impact, with high participation rate but no clear consensus on any item. There was division on whether FAccT should serve as a generator for alternative solidarity structures such as mutual aid (44\% agree, 20\% disagree, 36\% pass) and whether it should engage in real participatory design (48\% agree, 13\% disagree, 40\% pass). Similarly, assessments of FAccT's current performance were inconclusive: participants were split on whether the conference is falling short in providing tangible, practical guidance for non-academics to build responsible technology (41\% agree, 42\% pass) and whether it is not doing well in creating opportunities for participatory design, prototyping, and building together at the conference (38\% agree, 44\% pass). The high pass rates across these statements suggest considerable uncertainty about both FAccT's current practical impact and the extent to which it should expand in these directions. Overall, the community appears undecided on whether and how FAccT should move beyond academic exchange toward more hands-on, practice-oriented engagement.


\clearpage
\section{Additional Figures}
\begin{figure}[!ht]
    \centering
    \includegraphics[width=\textwidth]{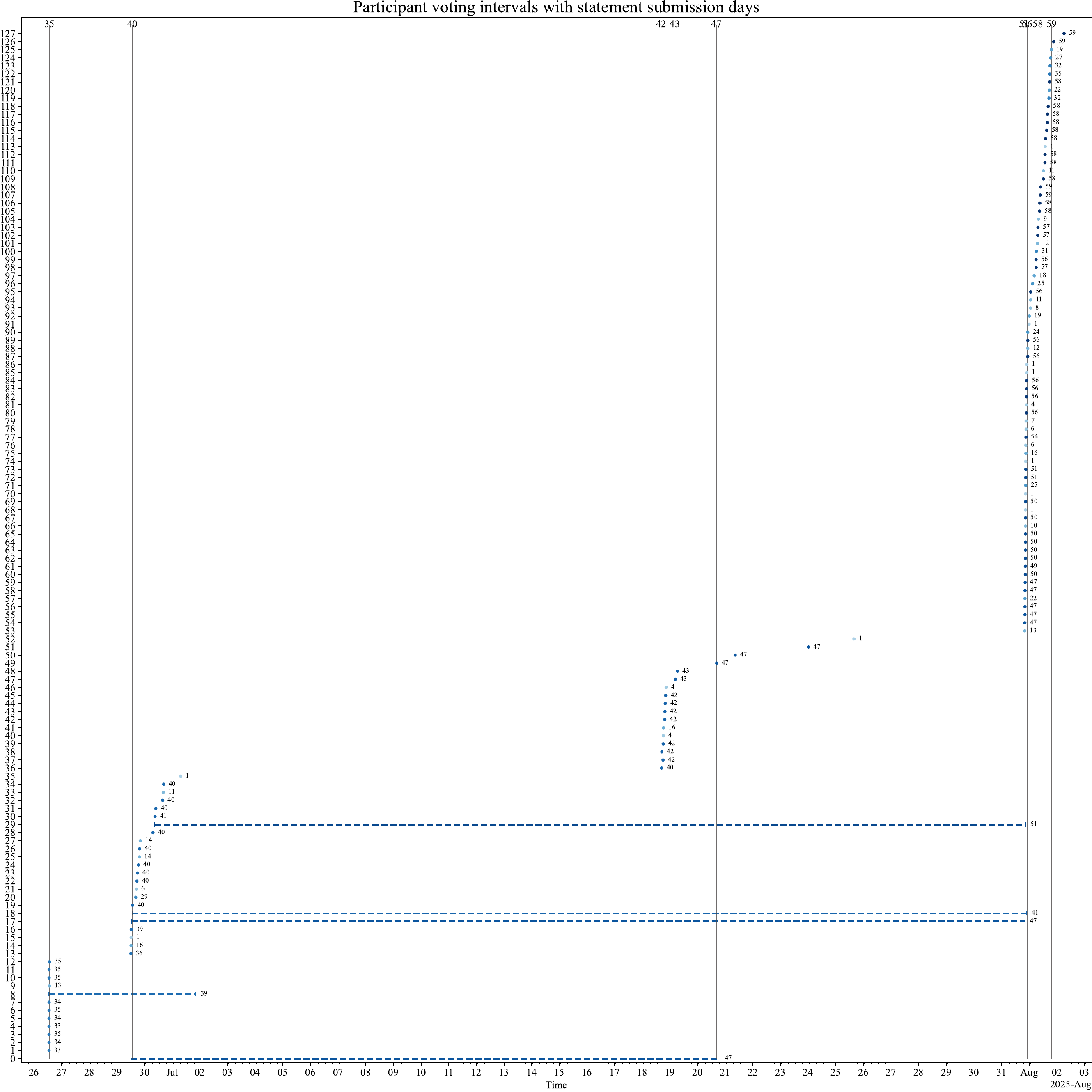}
    \caption{Participation intervals of all participants, represented by dashed line segment from the time of the first vote cast to the time of the last vote cast if the resulting interval was longer than 3 hours, else by single dot at the midpoint of the Participation interval. Vertical lines showed the change in the cumulative number of statements submitted, grouped by 6-hour windows. See also Fig.~\ref{fig:participation_interval}}
    \label{fig:participant_interval_full}
\end{figure}

\end{document}